\documentclass[fleqn,usenatbib]{aa}
\usepackage[varg]{txfonts}
\usepackage[dvipsnames]{xcolor} 
\usepackage{amsmath} 
\usepackage{booktabs}
\usepackage[utf8]{inputenc}
\usepackage{graphicx} 
\usepackage{xspace}
\usepackage{amsfonts}
\usepackage{ amssymb }
\usepackage{widetext}
\usepackage{ulem}

\newcommand{\ssout}[1]{}%
\newcommand{\add}[1]{{\color{black}{#1}}}
\newcommand{\FIXME}[1]{{\color{black}{#1}}}

\newcommand{\beq}{\begin{equation}}
\newcommand{\eeq}{\end{equation}}
\newcommand{\beqa}{\begin{eqnarray}}
\newcommand{\eeqa}{\end{eqnarray}}

\newcommand{\msun}{\ensuremath{M_{\odot}}\xspace}

\newcommand{\catalog}{GWTC-3\xspace}

\newcommand{\nn}{\nonumber}

\newcommand{\vl}{{\ensuremath{\vec{\lambda}} \xspace}}
\newcommand{\vd}{\ensuremath{D}\xspace}

\newcommand{\vt}{\ensuremath{\vec{\theta}} \xspace}
\newcommand{\ud}{\ensuremath{\mathrm{d}}}
\newcommand{\maHP}{\ensuremath{\mathcal{H}_{\mathrm{PE}}}\xspace}
\newcommand{\chieff}{\ensuremath{{\chi_{\rm{eff}}}}\xspace}
\newcommand{\gau}{\ensuremath{\mathfrak{g}}\xspace}
\newcommand{\bet}{\ensuremath{\mathfrak{b}}\xspace}
\newcommand{\ct}{\ensuremath{\cos{\tau}}\xspace}
\newcommand{\iso}{\ensuremath{\mathfrak{i}}\xspace}
\newcommand{\tuk}{\ensuremath{\mathfrak{t}}\xspace}

\newcommand{\Un}{\ensuremath{\mathcal{U}}}

\newcommand{\lvk}{\texttt{LVK default}\xspace}
\newcommand{\isoonly}{\texttt{Isotropic}\xspace}
\newcommand{\isobeta}{\texttt{Isotropic + Beta}\xspace}
\newcommand{\isogaus}{\texttt{Isotropic + Gaussian}\xspace}
\newcommand{\isotuk}{\texttt{Isotropic + Tukey}\xspace}
\newcommand{\isoggaus}{\texttt{Isotropic + 2 Gaussians}\xspace}
\newcommand{\isogbeta}{\texttt{Isotropic + Gaussian + Beta}\xspace}
\newcommand{\isogtuk}{\texttt{Isotropic + Gaussian + Tukey}\xspace}
\newcommand{\isocbeta}{\texttt{Isotropic + correlated Beta}\xspace}
\newcommand{\isocgaus}{\texttt{Isotropic + correlated Gaussian}\xspace}
\newcommand{\IsoGaussMuConstrained}{\FIXME{ \ensuremath{0.48^{+0.46}_{-0.99}}\xspace}}
\newcommand{\IsoGaussMuConstrainedSigmaLOne}{\FIXME{\ensuremath{0.41^{+0.47}_{-0.36}}\xspace}}
\newcommand{\IsoGaussMuConstrainedSigmaLDFive}{\FIXME{\ensuremath{0.28^{+0.35}_{-0.31}}\xspace}}
\newcommand{\IsoBetaMu}{\FIXME{\ensuremath{0.19^{+0.22}_{-0.18}}\xspace}}
\newcommand{\IsoTukeyMuTkLTwo}{\FIXME{\ensuremath{0.46_{-0.41}^{+0.44}}\xspace}}
\newcommand{\IsoTukeyMuTkLOne}{\FIXME{\ensuremath{0.30_{-0.27}^{+0.41}}\xspace}}
\newcommand{\IsoTukeyPercentFive}{\FIXME{\ensuremath{-0.37}}}
\newcommand{\IsoTukeyPercentNinetyFive}{\FIXME{\ensuremath{0.94}}}
\newcommand{\IsoGaussBetaMu}{\FIXME{\ensuremath{0.15_{-0.45}^{+0.39}}\xspace}}
\newcommand{\IsoTwoGaussMargA}{\FIXME{\ensuremath{-0.49_{-0.40}^{+0.42}}\xspace}}
\newcommand{\IsoTwoGaussMargSigmaLPercFive}{\FIXME{\ensuremath{0.46}}}
\newcommand{\IsoTwoGaussMargMuL}{\FIXME{\ensuremath{0.13_{-0.59}^{+0.52}}\xspace}}

\newcommand{\IsoGaussTukeyMu}{\FIXME{\ensuremath{0.34_{-1.1}^{+0.58}}\xspace}}
\newcommand{\IsoGaussTukeyMuTkLTwo}{\FIXME{\ensuremath{0.37_{-0.81}^{+0.53}}\xspace}}
\newcommand{\IsoGaussTukeyMuTkLOne}{\FIXME{\ensuremath{0.26_{-0.52}^{+0.54}}\xspace}}

\newcommand{\LIGOlabMIT}{LIGO Laboratory, Massachusetts Institute of Technology, 185 Albany St, Cambridge, MA 02139, USA}
\newcommand{\MKI}{Department of Physics and Kavli Institute for Astrophysics and Space Research, Massachusetts Institute of Technology, \\ 77 Massachusetts Ave, Cambridge, MA 02139, USA}

\graphicspath{{./}}

\begin{document}
\title{Spin it as you like: the (lack of a) measurement of the spin tilt distribution with LIGO-Virgo-KAGRA binary black holes}
\titlerunning{Spin it as you like}
\authorrunning{Vitale et al.}

\author{
Salvatore Vitale,$^{1,2}$\thanks{salvo@mit.edu}
Sylvia Biscoveanu,$^{1,2}$
Colm Talbot$^{1,2}$
}

\institute{
$^{1}$\LIGOlabMIT\\
$^{2}$\MKI
}

\abstract{The growing set of gravitational-wave sources is being used to measure the properties of the underlying astrophysical populations of compact objects, black holes and neutron stars. Most of the detected systems are black hole binaries. While much has been learned about black holes by analyzing the latest LIGO-Virgo-KAGRA (LVK) catalog, GWTC-3, a measurement of the astrophysical distribution of the black hole spin orientations remains elusive. This is usually probed by measuring the cosine of the tilt angle (\ct) between each black hole spin and the orbital angular momentum, $\ct=+1$ being perfect alignment.}{ \cite{LIGOScientific:2021psn} has modeled the \ct distribution as a mixture of an isotropic component and a Gaussian component {with mean} fixed at $+1$ and width measured from the data. We want to verify if the data \textit{require} the existence of such a peak at $\ct=+1$.} {We use various alternative models for the astrophysical tilt distribution and measure their parameters using the LVK GWTC-3 catalog.} {We find that a) Augmenting the LVK model such that the mean $\mu$ of the Gaussian is not fixed at $+1$ returns results that strongly depend on priors. If we allow $\mu>+1$ then the resulting astrophysical \ct distribution peaks at $+1$ and looks linear, rather than Gaussian. If we constrain $-1\leq \mu\leq+1$ the Gaussian component peaks at $\mu=\IsoGaussMuConstrained$ (median and 90\% symmetric credible interval). Two other 2-component mixture models yield \ct distributions that either have a broad peak centered at $\IsoBetaMu$ or a plateau that spans the range $[-0.5, +1]$, without a clear peak at $+1$. b) All of the models we considered agree on the fact that there is \textit{no} excess of black hole tilts at around $-1$. c) While yielding quite different posteriors, the models considered in this work have Bayesian evidences that are the same within error bars.} 
{We conclude that the current dataset is not sufficiently informative to draw any model-independent conclusions on the astrophysical distribution of spin tilts, except that there is no \textit{excess} of spins with negatively aligned tilts.}

\keywords{gravitational waves -- methods: data analysis  black hole physics}

\maketitle

\section{Introduction}

More than 90 binary black holes (BBHs) have been detected in the data of the ground-based gravitational-wave detectors LIGO~\citep{TheLIGOScientific:2014jea} and Virgo~\citep{TheVirgo:2014hva} by the LIGO-Virgo-Kagra (LVK) collaboration and other groups~\citep{LIGOScientific:2021djp, Nitz:2021zwj, Olsen:2022pin}. This dataset has been used to infer the properties of the underlying population---or populations---of BBHs. Among the parameters of interest, the masses and spins of the black holes play a prominent role since they can shed light on the binary formation channels\footnote{Eccentricity is also a powerful indicator of a binary formation channel~\citep{PhysRev.136.B1224, PhysRevD.77.081502, Morscher:2014doa, Samsing:2017xmd, Rodriguez:2017pec, Rodriguez:2018pss, Gondan:2018khr, Zevin:2017evb}, but it is currently harder to measure due to the limited sensitivity of ground-based detectors at frequencies below 20~Hz.}~\citep[e.g.,][]{Vitale:2015tea,Farr:2017uvj,Zevin:2017evb,Farr:2017gtv,Wong:2020ise, Zevin:2020gbd, Bouffanais:2021wcr}.

There currently exist a few approaches toward measuring the population properties. The first is to use a functional form for a reasonable population distribution, parameterized by some phenomenological parameters. For example, the LVK parameterized the primary mass distribution of the black holes as a mixture of a power law distribution and a Gaussian component~\citep{LIGOScientific:2021psn, Talbot:2018cva, Fishbach:2017zga}. This model includes several hyperparameters (since they pertain to the population as a whole, not to the individual events), which are measured from the data: the slope of the power law, the minimum and maximum black hole mass (including a smoothing parameter), the mean and standard deviation of the Gaussian component, and the branching ratio between the power law and Gaussian components. It is worth noting that not all parametric models make equally strong assumptions: an example of a more flexible parametrization is the Beta distribution that the LVK has used to describe the population distribution of black hole spin magnitudes~\citep{Wysocki:2018}. 
Those more elastic models might be more appropriate if one doesn't have strong observational or theoretical expectations about what the astrophysical distribution of a parameter should look like, or simply if they want to be more conservative. Just as Bayesian priors can significantly affect the posterior when the likelihood doesn't have a strong peak when analyzing individual compact binary coalescences, a hypermodel that is too strong could leave imprints on the inferred hyperparameters.

A second approach is to use non-parametric models, based on e.g. Gaussian processes~\citep{Tiwari:2020vym, Edelman:2021zkw, Rinaldi:2021bhm, Mandel:2016prl,Vitale:2018yhm}. Those usually have many more free parameters, which allows them to fit features in the data that parameterized models might miss. However, their larger number of parameters implies they might need more sources to reach a level of precision comparable to parametric models. 
Ideally, when strong parametric models are used, one would like to check that the results are not impacted by the model itself, and instead reveal features that are genuinely present in the data. A possible approach is to run multiple models (parametric and non parametric) and verify that they agree to within statistical uncertainties. 

Finally, recent work has focused on using as a model the predictions of numerical simulations. This typically involves applying machine learning~\citep{Wong:2020ise} or density estimation techniques~\citep{Zevin:2017evb, Bouffanais:2021wcr} to the binary parameter distributions output by rapid population synthesis codes or N-body simulations. While this approach is more astrophysically motivated, the population synthesis simulations have their own uncertainties and assumptions and often include so many free parameters that a complete exploration of the model space is not possible with current computational techniques~\citep[e.g.][]{Broekgaarden:2021efa}.
The impact of modeling on astrophysical inference of gravitational-wave sources has already become apparent in recent months. Several groups have investigated whether there is evidence for a fraction of black hole spin magnitude to be vanishingly small, finding results that depend on the model to a large extent~\citep{Callister:2022qwb,Galaudage:2021rkt,Tong:2022iws,Roulet:2021hcu,Mould:2022xeu} (Note that \cite{Callister:2022qwb} also provides a comprehensive summary of the status of that measurement).

In this paper we focus on the inference of the population distribution of tilt angles for the black hole binaries in the latest LVK catalog. This is the angle that each of the black hole spin vectors forms with the orbital angular momentum at some reference frequency (following the LVK data release, we will use $20$~Hz for the reference frequency\footnote{For the BBHs detected in the second half of the third observing run (O3b), the LVK has also released tilt posteriors evaluated at minus infinity, i.e. at very large orbital separations~\citep{Mould:2021xst}. We have run the \isobeta model of Sec.~\ref{sec.IsoPlusBeta} on O3b sources only, and found that the analyses with tilts calculated at $20$~Hz and minus infinity yield the same astrophysical \ct distribution. We also find that using O3b only sources the \ct distribution moves toward the left, compared to what is shown in Fig.~\ref{fig.IsoBeta_None_costau}, and peaks closer to 0.}). 
In their latest catalog of BBHs, the LVK collaboration has characterized this distribution by using a mixture model composed of an isotropic distribution plus a Gaussian distribution that peaks at $\ct=1$ (i.e. when the spin vector and the angular momentum are aligned) with unknown width~\citep{Talbot:2017yur, LIGOScientific:2021psn}. This model reflects expectations from astrophysical binary modeling. Indeed, numerical simulations suggest that binaries formed in galactic fields should have spins preferentially aligned with the angular momentum~\citep[e.g.,][]{1993MNRAS.260..675T, Kalogera:1999tq, Belczynski:2017gds, Zaldarriaga:2017qkw, Stevenson:2017tfq, Gerosa:2018wbw}, whereas binaries formed dynamically (i.e. in globular or star clusters) should have randomly oriented tilts~\citep[e.g.,][]{PortegiesZwart:2002iks, Rodriguez:2015oxa, Antonini:2016gqe, Rodriguez:2019huv, Gerosa:2021mno}. 

While this is a reasonable model, we are interested in verifying whether it is actually supported by the data in hand, or whether we are instead getting posteriors that are driven by that model. 
\cite{Callister:2022qwb} and \cite{Tong:2022iws} recently considered alternative models for the tilt angles, but they focused on whether there is a cutoff at negative cosine tilts (i.e. for anti-aligned spins), and if that answer depends on the model for the spin magnitude. On the other hand, we don't limit our investigation to the existence of negative tilts, but instead are interested in what---if anything---can be said about the tilt inference that is not strongly dependent on the model being used. 
We consider different alternative models and verify that all of them yield Bayesian evidences (and maximum log likelihood values) which are comparable with the default model used by the LVK. We also include models that allow for a correlation between \ct and other parameters (binary mass, mass ratio, or spin magnitude, in turn) and find that those too yield similar evidences. Critically, all of these alternative---and equally supported by the data---models yield noticeably different posterior distributions for the tilt population relative to the default LVK model. In particular, different models give different support to the existence and position of a feature at positive $\ct$. On the other hand, they all agree on the fact that there is no excess of systems with $\ct \simeq -1$. 
We conclude that the current constraints on the distribution of tilt angles are significantly affected by the model used, and that more sources (or weaker models) are needed before any conclusions can be drawn about the astrophysical distribution of binary black hole tilts. 

 \section{Methods}

We use hierarchical Bayesian analysis---extensively described in Appendix~\ref{Sec.Method}---to infer the astrophysical properties of the binary black holes reported in GWTC-3. 
To represent the astrophysical distribution of primary mass, mass ratio, redshift and spin magnitudes we use the flagship models used by the LVK in \cite{LIGOScientific:2021psn}. Namely, the primary mass distribution is their ``Power law $+$ Peak''~\citep{Talbot:2018cva}; the mass ratio is a power law~\citep{Fishbach:2019bbm}; the two spin magnitudes are independently and identically distributed according to a Beta distribution~\citep{Wysocki:2018}, and the redshift is evolving with a power law~\citep{Fishbach:2018edt}. For the (cosine\footnote{Even though we will only report results for the cosine of the tilt angle---\ct---we might occasionally refer to tilts only, to lighten the text.}) tilt distributions, we consider several models of increasing complexity.

\section{Results}\label{sec.Results}

In Appendix~\ref{Sec.Reference} we report our results when we use for the tilt distribution \add{the ``DEFAULT'' spin model of Appendix B-2-a} of \citet{LIGOScientific:2021psn} (referred to as \lvk in the rest of the paper): this will serve as a useful comparison for the more complex models described in the reminder of this work. In this section we re-analyze the GWTC-3 BBH with different parameterized two-component mixture models for \ct. We will report results for three-component models in Appendix~\ref{Sec.ThreeCompUncor}. It is assumed that the \ct distribution is not correlated to any other of the astrophysical parameters. This assumption will be revisited in Appendix~\ref{Sec.Correlated}.

\subsection{\isogaus model}~\label{sec.IsoPlusGaussian}

To check if the data \textit{requires} the that the Gaussian component of the \lvk mixture model peaks at $\ct=1$---Appendix~\ref{Sec.Reference}---we relax the assumption that the normal distribution must be centered at $+1$, That is, we treat the mean of the Gaussian component $\mu$ as another model parameter:
\beq
p(\cos\tau_1,\cos\tau_2 | \mu, \sigma,\gau) = \frac{1-\gau}{4} +\gau \prod_j^2{\mathcal{N}(\cos{\tau_j},\mu,\sigma)}
\label{Eq.IsoPlusGaussian}
\eeq
The prior for $\mu$ is uniform in the range $[-1,1]$ (Tab.~\ref{tab:priors} reports the priors for the hyperparameters of all models used in the paper).
\begin{figure}
\includegraphics[width=0.45\textwidth]{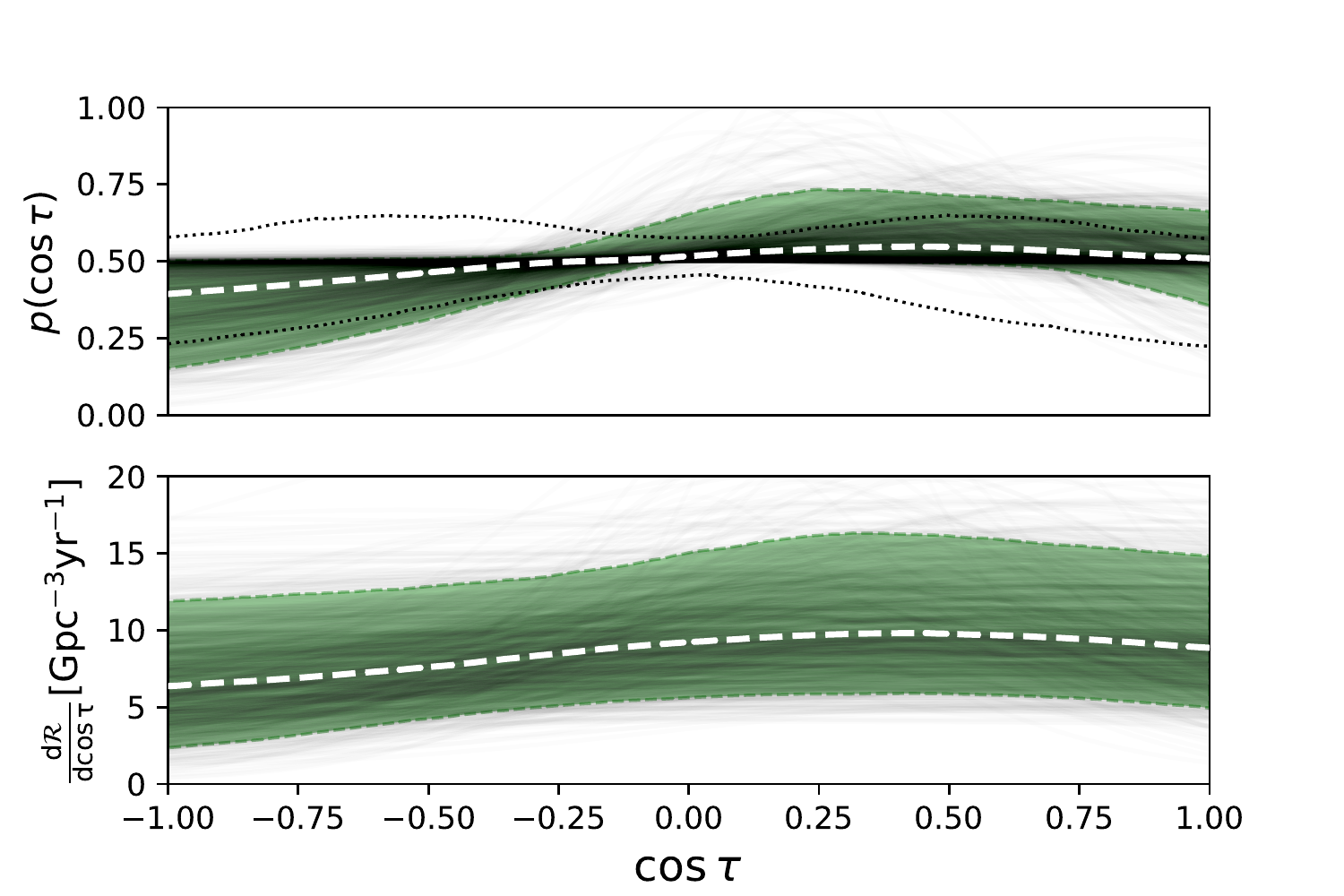}
\caption{(Top) Posterior for \ct obtained using the \isogaus model when the mean of the Gaussian component is allowed to vary in the range $\mu \in [-1,1]$. The two thin black dotted lines show 90\% credible interval obtained by drawing the model's hyperparameters from their priors. (Bottom) Differential merger rate per unit \ct for the same model. In both panels, the thin black lines represent individual posterior draws, whereas the colored band shows the 90\% credible interval. The thick dashed line within the band is the median.}
\label{fig.IsoMovingGaussian_None_ConstrainedMu_costau}
\end{figure}
The top panel of Fig.~\ref{fig.IsoMovingGaussian_None_ConstrainedMu_costau} shows the resulting posterior for the tilt angle. For the mean of the Gaussian component we measure $\mu=\IsoGaussMuConstrained$ (unless otherwise stated, we quote median and 90\% symmetric credible interval). Some of the uncertainty in this measurement is due to our choice to allow for large $\sigma$, since Gaussians with large $\sigma$ are rather flat, and can be centered anywhere without significantly affecting the likelihood. If we restrict the prior space to only allow for narrower Gaussians, then $\mu$ is much more constrained. E.g. if we limit to samples with $\sigma<0.5$ ($\sigma<1$) then $\mu=\IsoGaussMuConstrainedSigmaLDFive$ ($\mu=\IsoGaussMuConstrainedSigmaLOne$).
This can be also seen in a corner plot of the mean and standard deviation of the Gaussian component, Fig.~\ref{fig.IsoMovingGaussian_None_Combined_sns_mu_sigma} darker green. The data prefers positive means with standard deviations in the approximate range $\sigma \in [0.25,1.5]$. Smaller values of $\sigma$ are excluded, as are Gaussians centered at negative values of $\mu$, i.e. such that \chieff---the mass-weighted projection of the total spin along the angular momentum~\citep{Damour:2001}---would be negative. 
\add{The bottom panel of Fig.~\ref{fig.IsoMovingGaussian_None_ConstrainedMu_costau} shows the posterior on the differtial merger rate per unit \ct for the same model.  }

\begin{figure}
\centering
\includegraphics[width=0.5\textwidth]{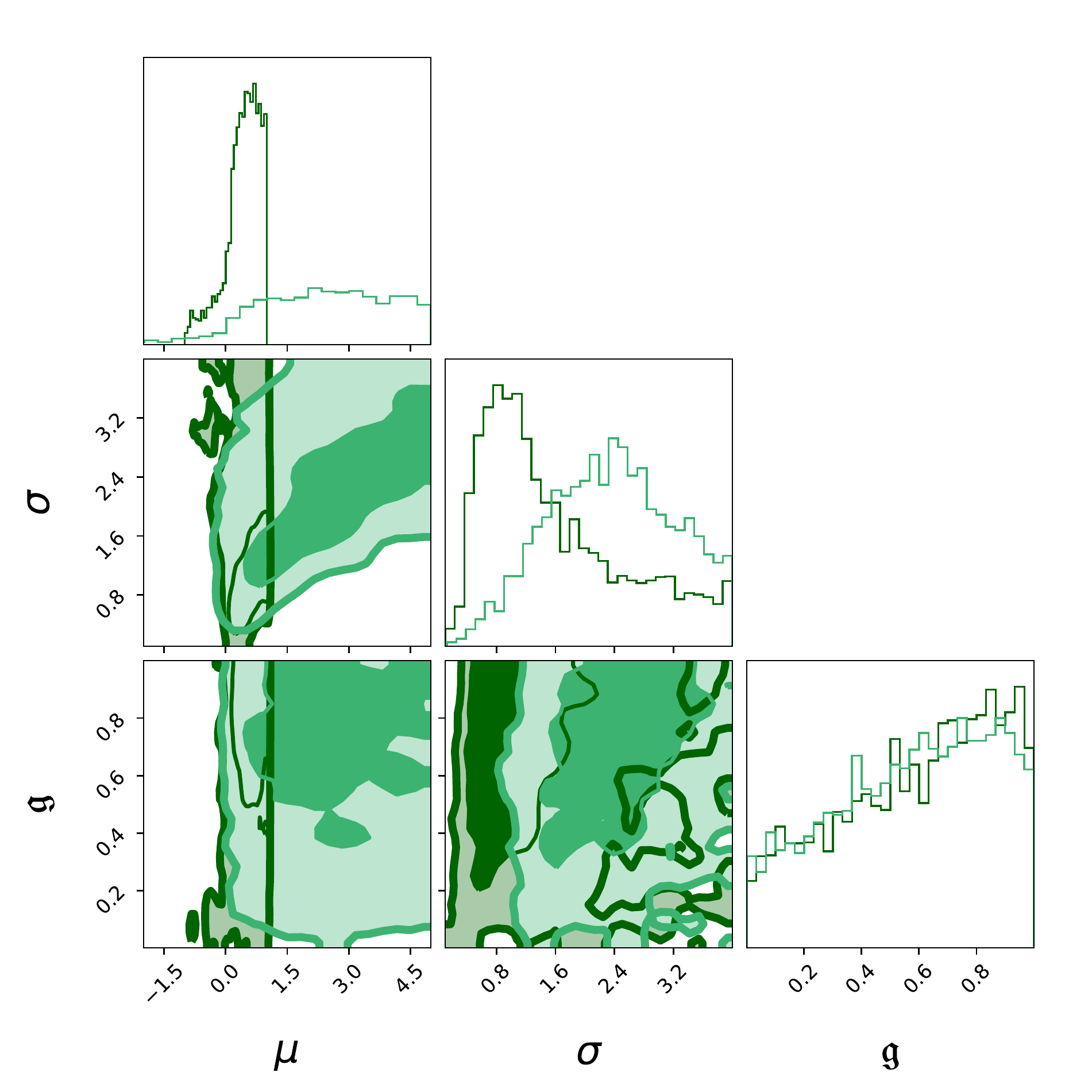}
\caption{Joint and marginal posteriors for the mean and standard deviation of the Gaussian component for the \isogaus model, as well as for the branching ratio \gau, when the mean of the Gaussian component is allowed to vary in the range $\mu \in [-1,1]$ (dark green) or $\mu \in [-5,5]$ (light green)}
\label{fig.IsoMovingGaussian_None_Combined_sns_mu_sigma}
\end{figure}

\begin{figure}
\centering
\includegraphics[width=0.5\textwidth]{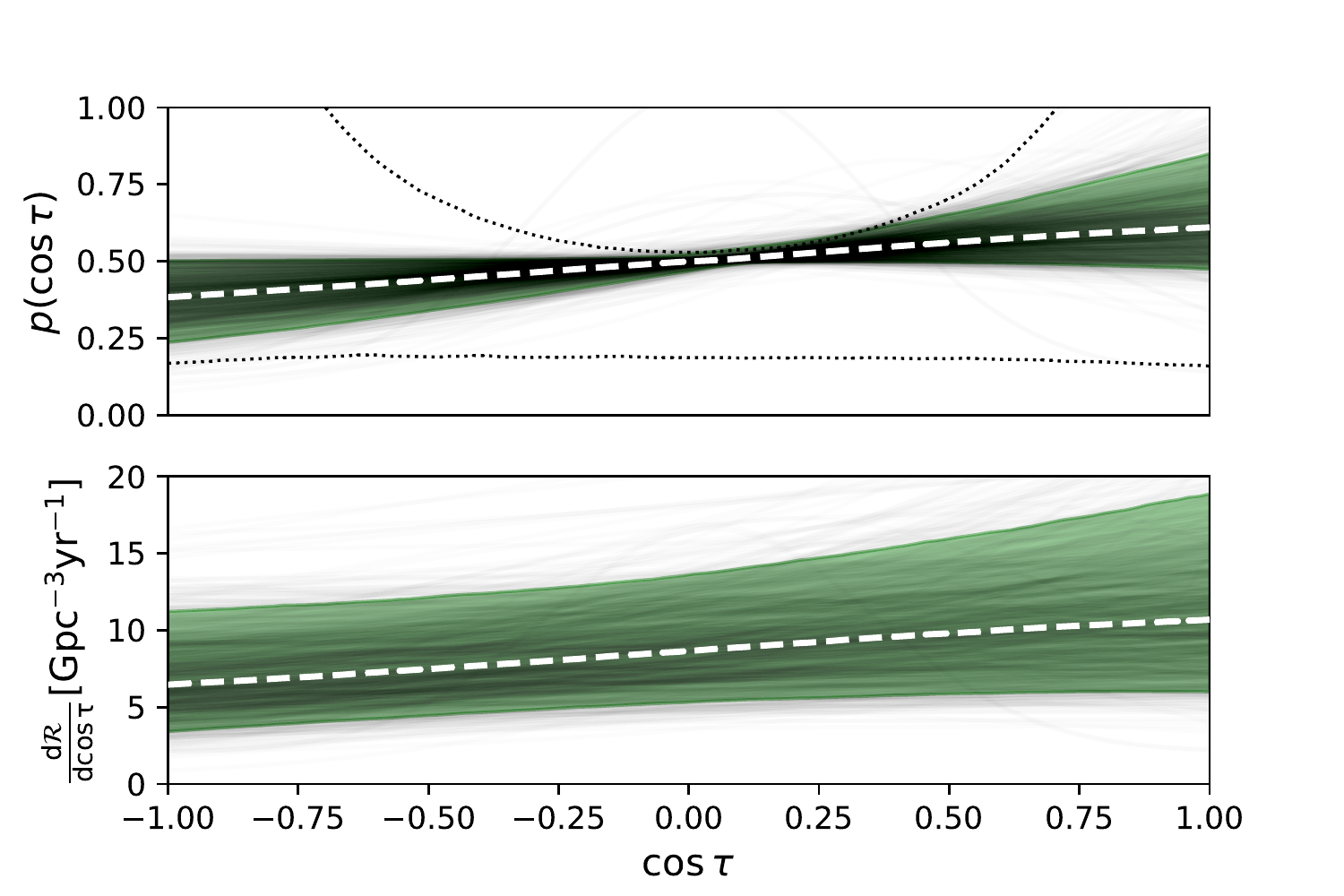}
\caption{Same as Fig.~\ref{fig.IsoMovingGaussian_None_ConstrainedMu_costau}, but for the \isogaus model, when the mean of the Gaussian component is allowed to vary in the range $\mu \in [-5,5]$.}
\label{fig.IsoMovingGaussian_None_UnconstrainedMu_costau}
\end{figure}

It is worth noting that the astrophysical \ct posterior changes entirely if the prior for $\mu$ is extended to allow for values outside of the range $[-1,1]$ (while the resulting population model is, of course, still truncated and properly normalized in that domain). 
For example, Fig.~\ref{fig.IsoMovingGaussian_None_UnconstrainedMu_costau} shows the posterior for the tilt distribution obtained with a  wider $\mu$ prior, uniform in the range $[-5,5]$. Here again it is the case that the distribution is consistent with having a peak for aligned spins. A look at the joint distribution of $\mu$ and $\sigma$, Fig.~\ref{fig.IsoMovingGaussian_None_Combined_sns_mu_sigma}, reveals that the peak at $\mu=1$ is not obtained because the Gaussian component peaks there, but rather by truncating Gaussians that peak at $\mu>1$ and have large standard deviations. This results in a much steeper shape for the posterior near the $\cos\tau=+1$ edge than what could possibly be obtained by forcing $-1\leq\mu\leq1$. The only feature that seems solid against model variations is the fact that there is no excess of systems at negative $\cos\tau$.

\begin{figure}
\includegraphics[width=0.5\textwidth]{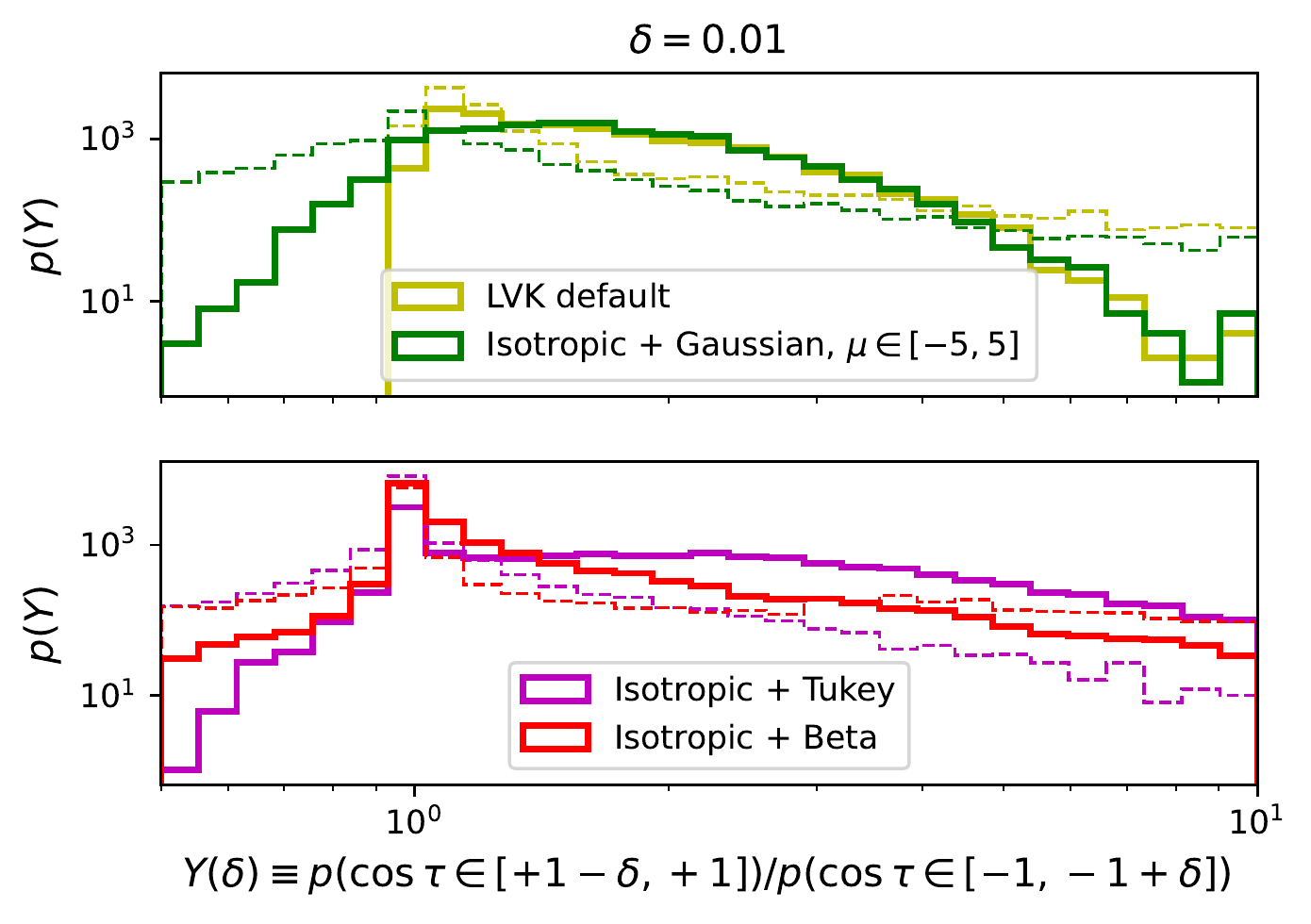}
\caption{An (unnormalized) histogram of $Y(\delta)$, the ratio between the probability of \ct for $\ct \in [1-\delta,1]$ and  $\ct \in [-1,-1+\delta]$ for the models of Sec.~\ref{sec.IsoPlusGaussian},\ref{sec.IsoPlusBeta} and \ref{sec.IsoPlusTukey} for $\delta=0.01$. We also show the result for the \lvk model (Appendix~\ref{Sec.Reference}) for comparison. The solid curves are obtained by sampling the hyperposteriors. The dashed line reports the same quantity, but drawing the model's hyperparameters from their priors. Values of $Y>1$ imply more support for aligned than anti-aligned spins.}
\label{fig.Ys_comparison}
\end{figure}

We can better visualize what happens at the edges of the \ct domain---i.e. for values of tilts close to aligned ($\ct \simeq 1$) or anti-aligned ($\ct \simeq -1$)---by plotting the ratio of the posterior support for aligned spin vs anti-aligned spins. This is equivalent to making a histogram of the ratio of the value that the thin black curves in Fig.~\ref{fig.IsoMovingGaussian_None_UnconstrainedMu_costau} take on the far right and far left side. Specifically, for each of the posterior draws, we calculate an asymmetry coefficient $Y$ defined as
\beq
Y(\delta) \equiv \frac{p(\ct \in [1-\delta,1])}{p(\ct \in [-1,-1+\delta])}\label{eq:y}
\eeq
and histogram it. When $Y=1$, the \ct distribution takes the same value at both of the edges; when $Y>1$ ($Y<1$) the \ct distribution has more support for aligned (anti-aligned) systems than for anti-aligned (aligned) ones. 
We notice that the \lvk model excludes \textit{a priori} excess of anti-aligned tilts, i.e. $Y<1$. This is instead not true for the other models we consider in this section. This is shown in the top panel of Fig.~\ref{fig.Ys_comparison}, for $\delta=0.01$ \add{(i.e. considering $\sim 8$ degree widths around $\pm \vec{L}$, where $ \vec{L}$ is the angular momentum vector)}. We see that the prior (dashed green curve) of $Y$ for the \isogaus model can extend to values smaller than $1$, i.e. can produce more anti-aligned spins than aligned ones (compare with the blue dashed curve and Appendix~\ref{Sec.Reference}). In fact, we see that the prior for this asymmetry probe is much less strong than in the default \lvk model as it does not exclude $Y<1$. As for the \lvk, the posterior of $Y$ is not inconsistent with $1$. Values of $Y$ smaller than $1$---i.e. an excess of anti-aligned spins---are severely suppressed relative to the prior, and so is a large excess of positive tilts. The posterior for $Y$ has a rather broad peak in the range $\sim [1,2]$ corresponding to distributions that are consistent with being either isotropic or having a mild excess of positive alignment.  In Appendix~\ref{App.AllYs} we show similar plots for different values of $\delta$.

\begin{figure}
\includegraphics[width=0.5\textwidth]{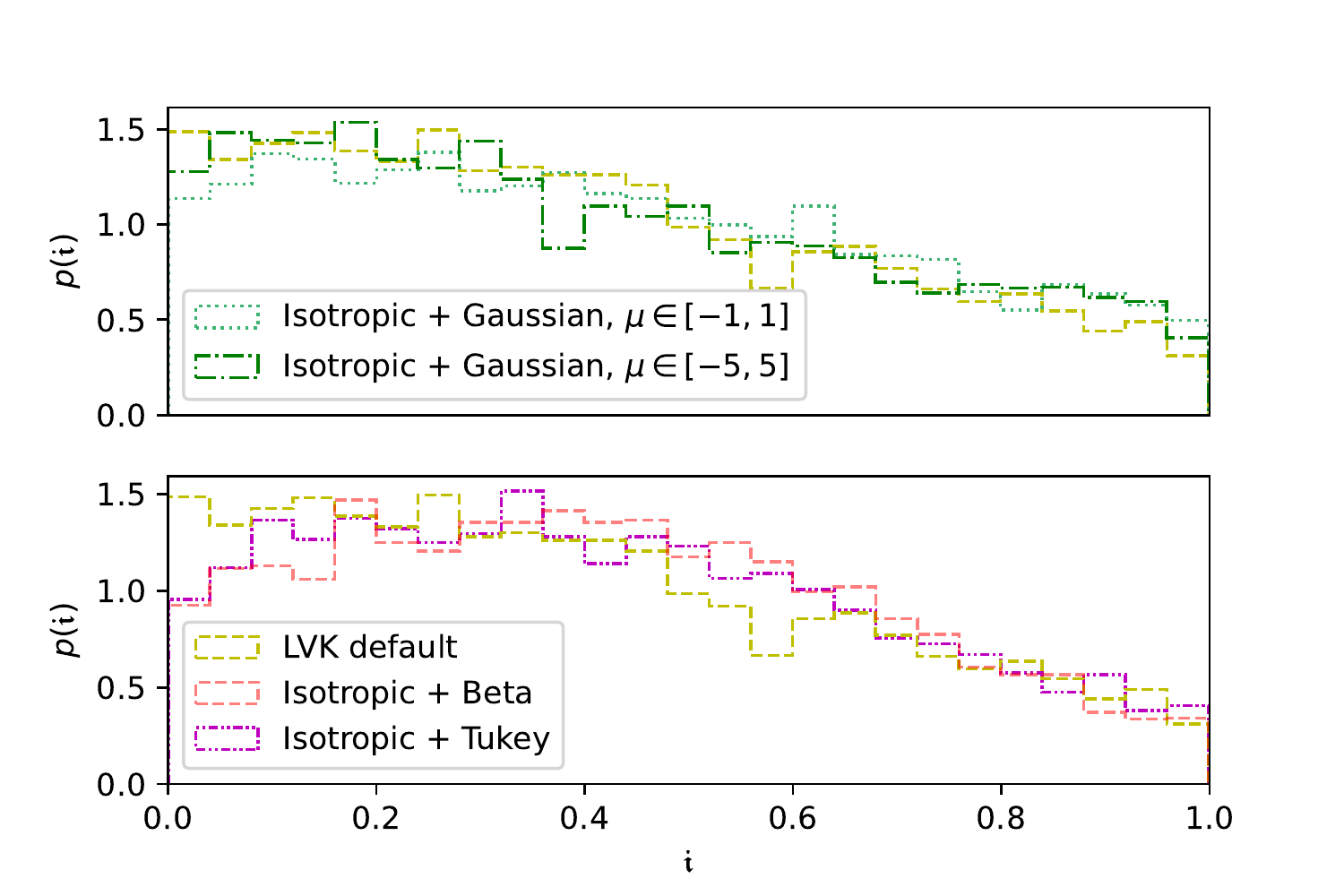}
\caption{Marginalized posterior for the branching ratio of the \textit{isotropic component} -- \iso -- for all of the uncorrelated two-component models, Sec.~\ref{sec.IsoPlusGaussian}, \ref{sec.IsoPlusBeta},~\ref{sec.IsoPlusTukey}. The figure is split into two panels to enhance clarity. In both panels, we also report the posterior obtained with the \lvk model (yellow dashed line) for comparison.}
\label{fig.iso_gaussian_models_pzeta}
\end{figure}

In the top panel of Figure~\ref{fig.iso_gaussian_models_pzeta} we show the marginalized posteriors of the branching ratio for the isotropic component, \iso, of the models described in this section, together with the reference \lvk model. While small variations exist, they all have support across the whole prior range, with a preference for small values of \iso.
\add{By comparing Figs.~\ref{fig.IsoMovingGaussian_None_ConstrainedMu_costau},~\ref{fig.IsoMovingGaussian_None_UnconstrainedMu_costau} and 
~\ref{fig.iso_gaussian_models_pzeta}, one may be surprised that the branching ratio posteriors for the two \isogaus runs are basically the same, and yet Fig.~\ref{fig.IsoMovingGaussian_None_ConstrainedMu_costau} seems to have a higher density of horizontal (i.e. isotropic) curves. This happens because the Gaussian component becomes isotropic for small $\mu$'s and large $\sigma$'s. Comparing the two distributions for $\mu$ on the top-left panel of Fig.~\ref{fig.IsoMovingGaussian_None_Combined_sns_mu_sigma} we see that indeed the run where $\mu$ is constrained to $[-1,1]$ has more support at small $\mu$'s, which, together with large $\sigma$'s, yield nearly horizontal posteriors in Fig.~\ref{fig.IsoMovingGaussian_None_ConstrainedMu_costau}. To a different extent the same is true for the models described below: there exist corners of the parameter space where the non-isotropic component can in fact generate a flat distribution, and unless otherwise said we won't \textit{a priori} exclude that possibility.}

\subsection{\isobeta model}~\label{sec.IsoPlusBeta}

To allow for a more elastic model for the non-isotropic component, we now replace the Gaussian component of the previous section with a Beta distribution:
\beq
p(\cos\tau_1,\cos\tau_2 | \alpha, \beta,\bet) = \frac{1-\bet}{4} +\bet \prod_j^2 \mathcal{B}(\cos \tau_j,\alpha,\beta)\label{Eq.IsoPlusBeta}
\eeq

\begin{figure}
\includegraphics[width=0.5\textwidth]{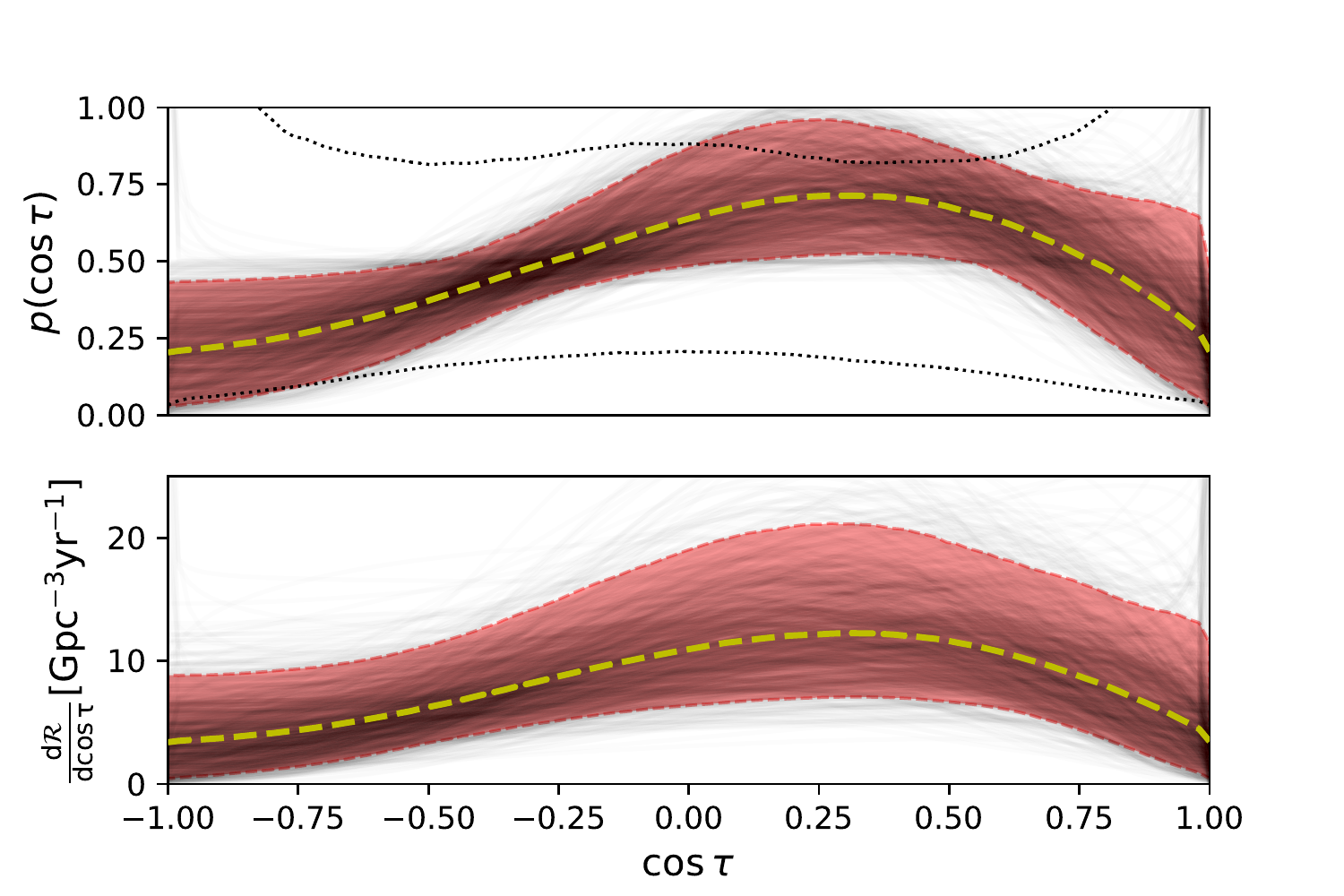}
\caption{Same as Fig.~\ref{fig.IsoMovingGaussian_None_ConstrainedMu_costau}, but for the \isobeta model. Note the different scale for the y axis of the bottom panel compared with similar figures for other models.}
\label{fig.IsoBeta_None_costau}
\end{figure}

We offset the input of the Beta distribution and scale its maximum value such that it spans the domain $[-1,1]$. We stress that we do \textit{not} limit the range of $\alpha$ and $\beta$ to non-singular values, i.e. we do allow them to be smaller than 1---Tab.~\ref{tab:priors}. In turn, this implies that we can get posteriors for \ct that peak at the edges of the range. The resulting posterior for $\cos\tau$ is shown in Fig.~\ref{fig.IsoBeta_None_costau}, which also shows for comparison the 90\% CI when drawing hyperparameters from their priors, thin dashed lines. With this model, we recover a broad peak at small positive values of \ct. One can convert the $\alpha$ and $\beta$ parameters of our rescaled Beta distribution to the corresponding mean as
\beq \mu_\beta= -1 + 2 \frac{\alpha}{\alpha+\beta}. \label{Eq.BetaMu}\eeq We find $\mu_\beta= \IsoBetaMu$. While some of the posterior draws do peak at $+1$, overall the upper edge of the 90\% CI band does not show a peak in that region. 
Compared to what seen in the previous section, this model finds more support for for anti-aligned tilts, with a median value that is at the lower edge of the 90\% CI for the \isogaus models. 
In the bottom panel of Fig.~\ref{fig.Ys_comparison} we show in red the posterior (solid line) and prior (dashed line) of the asymmetry coefficient $Y$ defined in Eq.~\ref{eq:y}. For this model too we observe that the posterior disfavours configurations with an excess of anti-aligned black hole tilts, relative to the prior. We notice that for the \isobeta model values of $Y>1$ do not necessarily represent a \textit{peak} at positive tilts, see Fig.~\ref{fig.IsoBeta_None_costau}, but only that negative tilts are even more suppressed. 
The marginalized posterior for the branching ratio of the isotropic component is shown in the bottom panel of Fig.~\ref{fig.iso_gaussian_models_pzeta} (histogram with dotted hatches). We find that, unlike the Gaussian-based models of Sec.~\ref{sec.IsoPlusGaussian} or the \lvk model, it does feature a very broad peak in the middle of the range. 

\subsection{\isotuk model}~\label{sec.IsoPlusTukey}

We end our exploration of two-component models with a mixture of an isotropic distribution and a distribution based on the Tukey window function. Mathematically:
\beqa
p(\cos\tau_1,\cos\tau_2 | \tuk,T_{x0},T_k,T_r) = \frac{1-\tuk}{4} && \nonumber \\ 
+\tuk \prod_j^2 {\mathcal{T}(\cos{\tau_j},T_{x0},T_k,T_r)}
\label{Eq.IsoPlusTukey}
\eeqa
where the exact expression for the functional form of the window function, and a few examples are given in Appendix~\ref{Sec.AppTukey}. Figure~\ref{fig.IsoPlusTukeyExt_None_costau} reports the resulting posterior distribution for \ct, together with the prior (dotted black lines). The 90\% CI band features a plateau that extends from $\ct \simeq -0.5$ to $+1$. 

\begin{figure}
\includegraphics[width=0.5\textwidth]{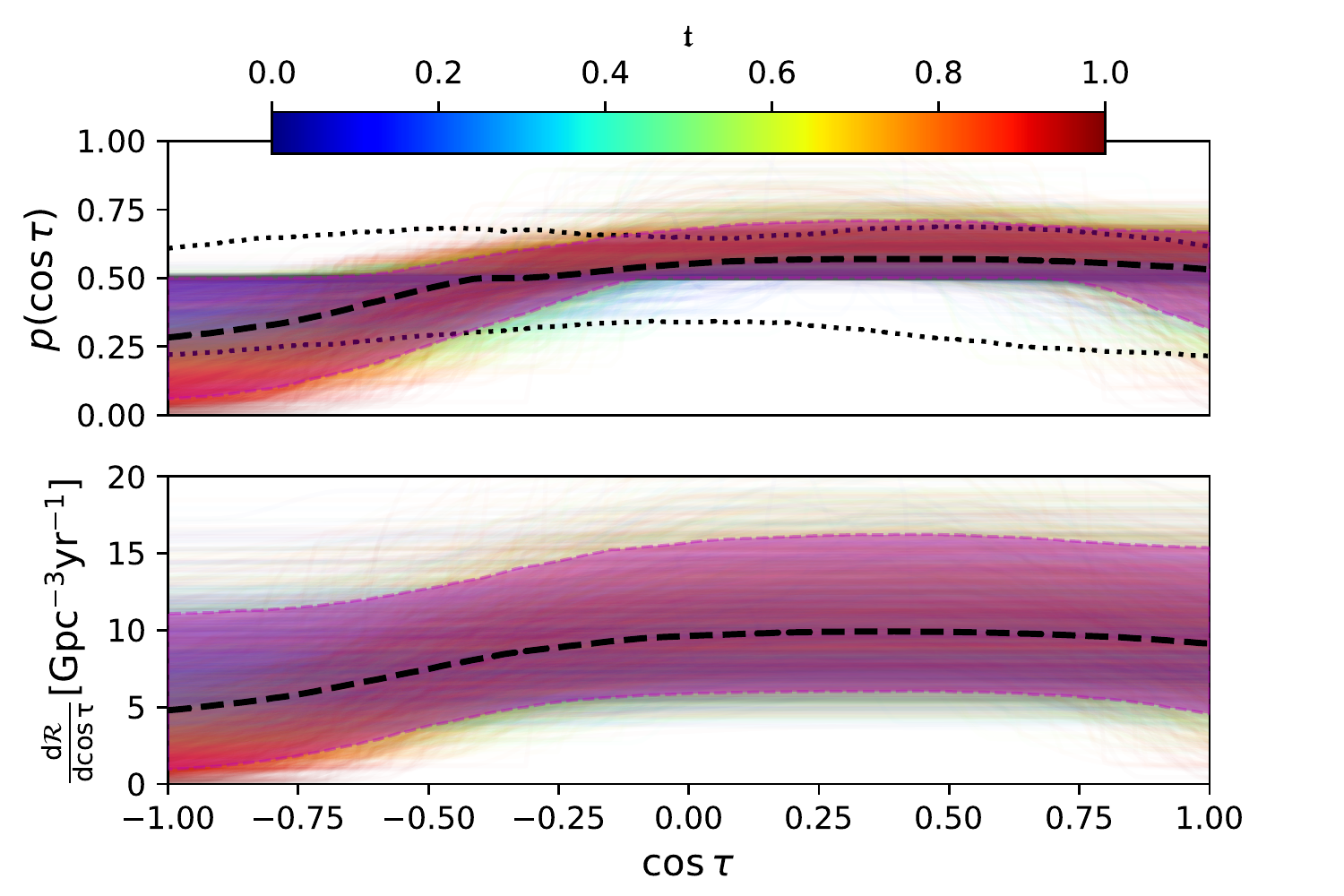}
\caption{ Same as Fig.~\ref{fig.IsoMovingGaussian_None_ConstrainedMu_costau} bur for the \isotuk model. Individual posterior draws are colored according to the corresponding value of the branching ratio for the Tukey component, \tuk.}
\label{fig.IsoPlusTukeyExt_None_costau}
\end{figure}

\begin{figure}
\includegraphics[width=0.45\textwidth]{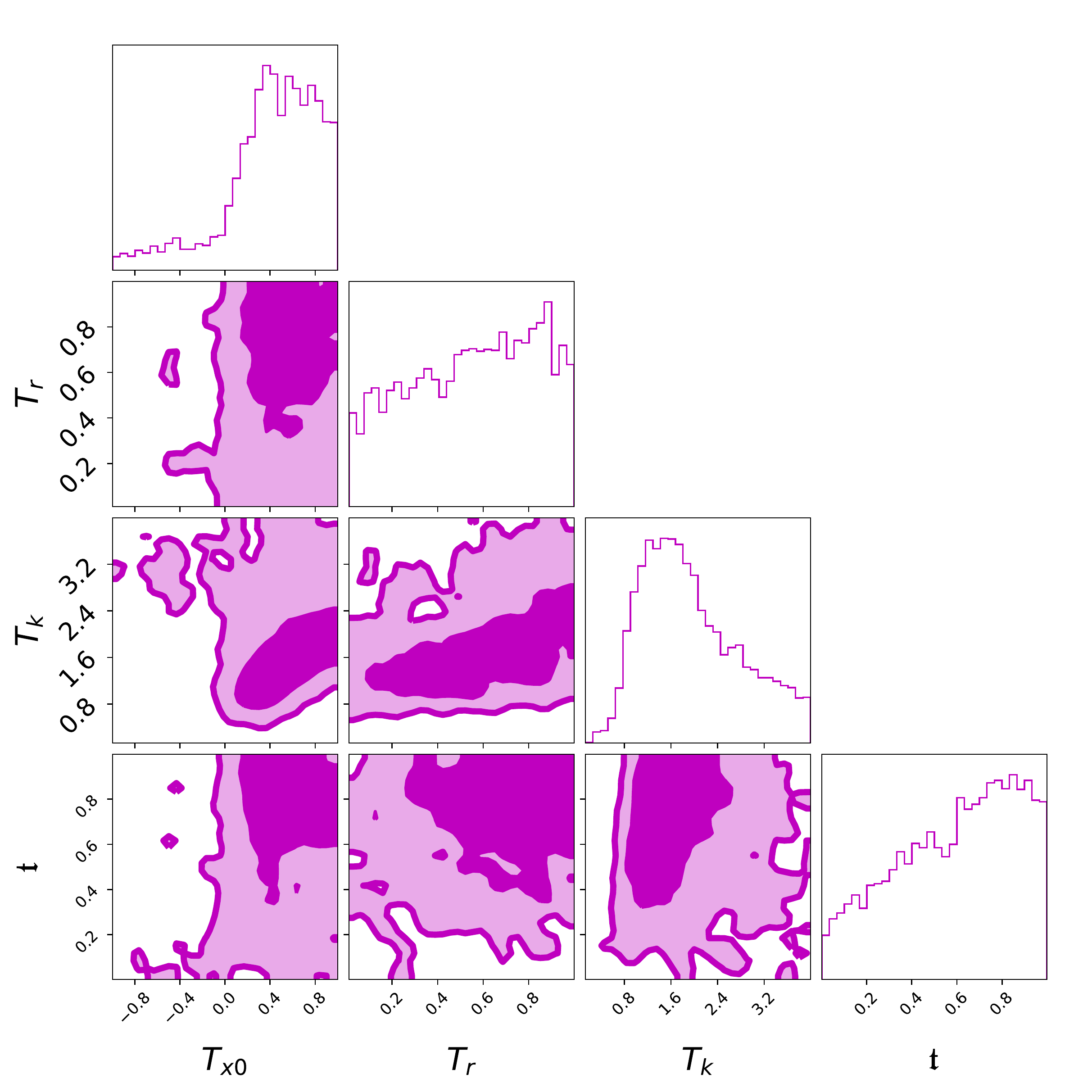}
\caption{Joint and marginal posteriors for the hyperparameters and branching ratio of the Tukey component in the \isotuk model.}
\label{fig.IsoPlusTukeyExt_None_TukeyPars}
\end{figure}

In Fig.~\ref{fig.IsoPlusTukeyExt_None_TukeyPars}, we show the posterior distribution for the parameters controlling the Tukey channel, together with the corresponding branching ratio. The Tukey component is  centered at $T_{x0}$, whose $5$th and $95$th percentile are $\IsoTukeyPercentFive$ and $\IsoTukeyPercentNinetyFive $ respectively. 
The marginal posterior for $T_k$ prefers values close to $1.8$, implying wider Tukey distributions. Smaller values of $T_k$ are possible only for small \tuk as expected given that for small \tuk the data cannot constrain the Tukey component, and the posterior must then resemble the uniform prior, which includes small $T_k$. The posteriors of these two parameters are correlated such that when $\tuk$ is large, $T_k$ is also large, meaning the resulting \ct distribution more closely resembles an isotropic distribution. However, when $T_k$ gets larger than $\sim 2$ then \tuk is not constrained at all. This happens because when $T_k$ is that large the Tukey component is extremely close to an isotropic distribution, at which point the whole model is isotropic, and the branching ratio stops being a meaningful parameter. Finally, the posterior for $T_r$ is wide, with a preference for larger values, implying a Tukey distribution that ramps up and down smoothly rather than producing sharp features.  
Figure~\ref{fig.IsoPlusTukeyExt_None_TukeyPars} also reveals that large values of $T_k$ are responsible for the near entirety of the support at $T_{x0}<0$, since a Tukey that is basically a uniform distribution can be centered anywhere without affecting the likelihood. If we restrict the analysis to $T_k\leq 2$ ($T_k\leq 1$), we get that $T_{x0}$ is much better constrained to  $T_{x0}= \IsoTukeyMuTkLTwo$ ($T_{x0}=\IsoTukeyMuTkLOne$), which excludes negative values at $\sim$90\% credibility.
The fact that our generous hyperparameter priors allow for Tukey distributions that resemble isotropic ones also explain the peak at $Y=1$ in the bottom panel of Fig.~\ref{fig.Ys_comparison}, purple lines (solid for the posterior, dashed for the prior). We find, once again, that the data does not exclude that the \ct distribution is in fact isotropic, and the only solid conclusion one can make seems to be that there is no excess of systems at $\ct \simeq -1$, since $p(Y)$ is heavily suppressed relative to its prior for $Y\lesssim 1$.

\section{Conclusions}

In this paper we have re-analyzed the LVK's 69 BBHs of GWTC-3 using different models for the astrophysical distribution of the black hole spin tilt angle, i.e. the angle the spin vector forms with the orbital angular momentum at a reference frequency ($20$~Hz).
Black hole spin tilts can yield precious information about their astrophysical formation channels. It is usually expected that dynamical formation of binaries results in an isotropic distribution of the spin vectors~\citep[e.g.,][]{PortegiesZwart:2002iks, Rodriguez:2015oxa, Antonini:2016gqe, Rodriguez:2019huv, Gerosa:2021mno}. On the other hand, for black hole binaries formed in the field via isolated binary evolution, it is expected that the spins are nearly aligned with the angular momentum~\citep[e.g.,][]{1993MNRAS.260..675T, Kalogera:1999tq, Belczynski:2017gds, Zaldarriaga:2017qkw, Stevenson:2017tfq, Gerosa:2018wbw}, i.e. that tilts are small, as the angular momenta of the progenitor stars are aligned by star-star and star-disk interactions~\citep{1981A&A....99..126H, 1981A&A...102...17P}. 
Indeed, if the binary forms in the field, the only mechanism that could yield significant black hole tilts are asymmetries in the supernovae explosions that create the black holes. These asymmetries can impart a natal kick large enough to tilt the orbital plane~\citep{1975Natur.253..698K, Kalogera:1999tq, Hurley:2002rf}. However, the black hole natal kick distribution is poorly understood both theoretically~\citep[e.g.,][]{Dominik:2012kk, Zevin:2017evb, Mapelli:2018wys, Repetto:2012bs, Giacobbo:2019fmo, Fragos:2010tm} and observationally~\citep{1995MNRAS.277L..35B, 1999A&A...352L..87N, 2001Natur.413..139M, 2002A&A...395..595M, Wong:2013vya}.

These expectations explain why in their most recent catalog, the LVK has modeled the astrophysical tilt distribution as a mixture of two components: an isotropic part and a Gaussian distribution centered at $\ct=1$ and with a width that is measured from the data. This is a rather strong model, as it forces onto the data a Gaussian that \textit{must} be centered at $+1$. This might not be advisable, because individual tilt measurements are usually broad, which implies that the functional form of the assumed astrophysical model can leave a discernible imprint on the posterior. 
Given the rather large uncertainties about how much spin misalignment can be produced in each channel, and the fact that the BBH population currently on hand might contain back holes formed in different channels~\citep{Zevin:2020gbd, Wong:2020ise, Bouffanais:2021wcr, Franciolini:2022iaa}, it is legitimate to question whether the data \textit{requires} that (as opposed to: is consistent with) the \ct distribution peaks at $+1$. We find that it does not. 

We consider {three} 2-component mixture models: \isogaus, made of an isotropic component and a Gaussian component whose mean is \textit{not} fixed at $+1$, but rather measured from the data; \isobeta, made of an isotropic component and a (singular) Beta component; \isotuk, made of an isotropic component and a distribution based on the Tukey window. 
We find that the only model with a posterior that peaks at $\ct=1$ is the \isogaus, and only if we allow the mean of the Gaussian to take values outside of the \ct domain $[-1,1]$: Fig.~\ref{fig.IsoMovingGaussian_None_Combined_sns_mu_sigma} shows that in this case the Gaussian component peaks at $\mu>1$ and is very broad, yielding a peak at $+1$, Fig.~\ref{fig.IsoMovingGaussian_None_UnconstrainedMu_costau}.
The other two models yield either a posterior that peaks at $\ct \simeq 0.2$ and no peak at $+1$, or a plateau from $\ct \simeq -0.5$ to $+1$. For all of the above models the data is not decisively ruling out a fully isotropic tilt distribution, but is \textit{inconsistent} with an excess of systems with large and negative tilts. \add{The models we considered found features---on top of an isotropic distribution---whose exact shape depends to a large extent on the model's flexibility. Therefore, the fact that the \lvk model finds support for a peak at $\ct=1$ can be explained because it can only \textit{add} support at $\ct=1$ when trying to match any population features on top of the isotropic distribution. }
Our results agree with previous results for population models fitting the distribution of \chieff, which find only a small fraction of sources with negative \chieff, implying negative tilts~\citep{Roulet:2018jbe, Miller:2020zox, Roulet:2021hcu, Callister:2022qwb}. Indeed, if we recast our inferred distributions for \ct and spin magnitude to the resulting \chieff distribution, we obtain results consistent with \cite{LIGOScientific:2021psn}. 
In the appendices below we report on other models. In Appendix ~\ref{Sec.ThreeCompUncor} we  consider {three} 3-component mixture models: \isoggaus, made of an isotropic component and 2 Gaussian components;  \isogbeta, made of an isotropic component a (non-singular) Beta component and a Gaussian component; \isogtuk, made of an isotropic component, a distribution based on the Tukey window and a Gaussian component. These more elastic models are consistent with a broad plateau in the \ct posterior that extends from $\sim-0.5$ to $+1$. Whether there also are peaks or features at small positive \ct and/or at $+1$ depends on the exact model. 

In Appendix ~\ref{Sec.Correlated} we augment some of the 2-component mixture models to allow for correlations between the tilt angles and another of the binary parameters: component masses, component spins, mass ratio and {total mass} in turn. One might expect some correlation as the mechanisms that can misalign the binary orbital plane or the black hole spins are affected by the binary parameters~\citep[e.g.][]{1994A&A...290..496J, Burrows:1995bb, Fryer:2005sz, Gerosa:2021mno}. To keep the number of model parameters limited---consistently with the relatively limited number of sources---we only consider linear correlations, and only correlate the tilt distribution with one other parameter at a time. However, even with these limitations we find that the current dataset cannot significantly constrain eventual correlations.  For all of the models considered in this work, we report Bayesian evidences, Tab.~\ref{tab:Evidences}, which might be used to calculate odds ratios. We find that---within sampling and numerical uncertainties---all of the models are equally supported by the data. 

We conclude that the current dataset is not yet large and informative enough to prove that the astrophysical tilt distribution has features, nor that {if} features exist they manifest as an excess of systems with nearly aligned spin vectors. On the contrary, most of the models we considered yield a broad peak in the astrophysical \ct distribution at small and positive values. The only conclusion that is consistently found across all models is that there is no excess of systems with negative tilts, relative to what is expected in an isotropic distribution. Our results agree with the literature~\citep[e.g.][]{Callister:2022qwb,Tong:2022iws, LIGOScientific:2021psn,Mould:2022xeu} on the lack of an excess of $\ct \simeq -1$ but disagree on other details (e.g. whether there is an hard cutoff in the \ct distribution at $\ct<0$ \citep[cfr][]{Callister:2022qwb}). The point of this work is to show that those disagreements are to be expected, given the information in the current dataset. The next observing run of LIGO, Virgo and KAGRA is scheduled to start in early 2023~\citep{KAGRA:2013rdx} and should yield hundreds of BBH sources. Those may yield a first firm measurement of the astrophysical distribution of the tilt angle, and possibly allow us to begin probing correlations with other astrophysical parameters. 

\section*{Acknowledgments}

The authors would like to thank C. Adamcewicz, V. Baibhav, T. Dent, S. Galaudage, C. Rodriguez and M. Zevin for useful comments and discussion.
We would in particular like to thank T. Callister and D. Gerosa for many insightful comments and suggestions. We would like to thank the anonymous A\&A Referee, whose comments helped improve the manuscript. 
S.V. is supported by NSF through the award PHY-2045740. 
S.B. is supported by the NSF Graduate Research Fellowship under Grant No. DGE-1122374.
CT is supported by the MKI Kavli Fellowship.
This material is based upon work supported by NSF's LIGO Laboratory which is a major facility fully funded by the National Science Foundation.
This paper carries LIGO document number LIGO-P2200275.

\section*{Data Availability}

A public repository with the hyperposteriors used in this work will be available on \href{https://doi.org/10.5281/zenodo.7297514
}{Zenodo}. We used publicly-available programs \textsc{Bilby}~\citep{Ashton:2018jfp, Romero-Shaw:2020owr}, \textsc{dynesty}~\citep{Speagle_2020} and \textsc{GWPopulation}~\citep{Talbot:2019okv}

\bibliographystyle{aa}
\bibliography{tilt} 

\begin{thebibliography}{89}
\expandafter\ifx\csname natexlab\endcsname\relax\def\natexlab#1{#1}\fi

\bibitem[{Aasi {et~al.}(2015)}]{TheLIGOScientific:2014jea}
Aasi, J. {et~al.} 2015, Class. Quant. Grav., 32, 074001

\bibitem[{Abbott {et~al.}(2018)}]{KAGRA:2013rdx}
Abbott, B.~P. {et~al.} 2018, Living Rev. Rel., 21, 3

\bibitem[{Abbott {et~al.}(2020{\natexlab{a}})}]{GWTC2Release}
Abbott, R. {et~al.} 2020{\natexlab{a}},
  https://dcc.ligo.org/LIGO-P2000223/public

\bibitem[{Abbott {et~al.}(2020{\natexlab{b}})}]{GWTC1Release}
Abbott, R. {et~al.} 2020{\natexlab{b}},
  https://dcc.ligo.org/LIGO-P1800370/public

\bibitem[{Abbott
  {et~al.}(2021{\natexlab{a}})}]{ligo_scientific_collaboration_and_virgo_2021_5117703}
Abbott, R. {et~al.} 2021{\natexlab{a}}, {GWTC-2.1: Deep Extended Catalog of
  Compact Binary Coalescences Observed by LIGO and Virgo During the First Half
  of the Third Observing Run - Parameter Estimation Data Release}

\bibitem[{Abbott {et~al.}(2021{\natexlab{b}})}]{LIGOScientific:2021djp}
Abbott, R. {et~al.} 2021{\natexlab{b}}, arXiv e-prints, arXiv:2111.03606

\bibitem[{Abbott
  {et~al.}(2021{\natexlab{c}})}]{ligo_scientific_collaboration_and_virgo_2021_5546676}
Abbott, R. {et~al.} 2021{\natexlab{c}}, {GWTC-3: Compact Binary Coalescences
  Observed by LIGO and Virgo During the Second Part of the Third Observing Run
  — O3 search sensitivity estimates}

\bibitem[{Abbott
  {et~al.}(2021{\natexlab{d}})}]{ligo_scientific_collaboration_and_virgo_2021_5546663}
Abbott, R. {et~al.} 2021{\natexlab{d}}, {GWTC-3: Compact Binary Coalescences
  Observed by LIGO and Virgo During the Second Part of the Third Observing Run
  — Parameter estimation data release}

\bibitem[{Abbott {et~al.}(2021{\natexlab{e}})}]{LIGOScientific:2021psn}
Abbott, R. {et~al.} 2021{\natexlab{e}}, arXiv e-prints, arXiv:2111.03634

\bibitem[{Acernese {et~al.}(2015)}]{TheVirgo:2014hva}
Acernese, F. {et~al.} 2015, Class. Quant. Grav., 32, 024001

\bibitem[{Adamcewicz \& Thrane(2022)}]{Adamcewicz:2022hce}
Adamcewicz, C. \& Thrane, E. 2022, Arxiv preprints [\eprint[arXiv]{2208.03405}]

\bibitem[{Antonini \& Rasio(2016)}]{Antonini:2016gqe}
Antonini, F. \& Rasio, F.~A. 2016, Astrophys. J., 831, 187

\bibitem[{Ashton {et~al.}(2019)}]{Ashton:2018jfp}
Ashton, G. {et~al.} 2019, Astrophys. J. Suppl., 241, 27

\bibitem[{Belczynski {et~al.}(2020)}]{Belczynski:2017gds}
Belczynski, K. {et~al.} 2020, Astron. Astrophys., 636, A104

\bibitem[{Biscoveanu {et~al.}(2022)Biscoveanu, Callister, Haster, Ng, Vitale,
  \& Farr}]{Biscoveanu:2022qac}
Biscoveanu, S., Callister, T.~A., Haster, C.-J., {et~al.} 2022, Astrophys. J.
  Lett., 932, L19

\bibitem[{Biscoveanu {et~al.}(2021)Biscoveanu, Isi, Vitale, \&
  Varma}]{Biscoveanu:2020are}
Biscoveanu, S., Isi, M., Vitale, S., \& Varma, V. 2021, Phys. Rev. Lett., 126,
  171103

\bibitem[{Bouffanais {et~al.}(2021)Bouffanais, Mapelli, Santoliquido, Giacobbo,
  Di~Carlo, Rastello, Artale, \& Iorio}]{Bouffanais:2021wcr}
Bouffanais, Y., Mapelli, M., Santoliquido, F., {et~al.} 2021, Mon. Not. Roy.
  Astron. Soc., 507, 5224

\bibitem[{{Brandt} {et~al.}(1995){Brandt}, {Podsiadlowski}, \&
  {Sigurdsson}}]{1995MNRAS.277L..35B}
{Brandt}, W.~N., {Podsiadlowski}, P., \& {Sigurdsson}, S. 1995, \mnras, 277,
  L35

\bibitem[{Broekgaarden {et~al.}(2021)}]{Broekgaarden:2021efa}
Broekgaarden, F.~S. {et~al.} 2021, Mon. Not. Roy. Astron. Soc.
  [\eprint[arXiv]{2112.05763}]

\bibitem[{Burrows \& Hayes(1996)}]{Burrows:1995bb}
Burrows, A. \& Hayes, J. 1996, Phys. Rev. Lett., 76, 352

\bibitem[{Callister {et~al.}(2021)Callister, Haster, Ng, Vitale, \&
  Farr}]{Callister:2021fpo}
Callister, T.~A., Haster, C.-J., Ng, K. K.~Y., Vitale, S., \& Farr, W.~M. 2021,
  Astrophys. J. Lett., 922, L5

\bibitem[{{Callister} {et~al.}(2022){Callister}, {Miller}, {Chatziioannou}, \&
  {Farr}}]{Callister:2022qwb}
{Callister}, T.~A., {Miller}, S.~J., {Chatziioannou}, K., \& {Farr}, W.~M.
  2022, arXiv e-prints, arXiv:2205.08574

\bibitem[{Damour(2001)}]{Damour:2001}
Damour, T. 2001, Phys. Rev. D, 64, 124013

\bibitem[{Dominik {et~al.}(2012)Dominik, Belczynski, Fryer, Holz, Berti, Bulik,
  Mandel, \& O'Shaughnessy}]{Dominik:2012kk}
Dominik, M., Belczynski, K., Fryer, C., {et~al.} 2012, Astrophys. J., 759, 52

\bibitem[{Edelman {et~al.}(2022)Edelman, Doctor, Godfrey, \&
  Farr}]{Edelman:2021zkw}
Edelman, B., Doctor, Z., Godfrey, J., \& Farr, B. 2022, Astrophys. J., 924, 101

\bibitem[{Farr {et~al.}(2018)Farr, Holz, \& Farr}]{Farr:2017gtv}
Farr, B., Holz, D.~E., \& Farr, W.~M. 2018, Astrophys. J. Lett., 854, L9

\bibitem[{{Farr}(2019)}]{2019RNAAS...3...66F}
{Farr}, W.~M. 2019, Research Notes of the American Astronomical Society, 3, 66

\bibitem[{Farr {et~al.}(2017)Farr, Stevenson, Coleman~Miller, Mandel, Farr, \&
  Vecchio}]{Farr:2017uvj}
Farr, W.~M., Stevenson, S., Coleman~Miller, M., {et~al.} 2017, Nature, 548, 426

\bibitem[{Fishbach \& Holz(2017)}]{Fishbach:2017zga}
Fishbach, M. \& Holz, D.~E. 2017, Astrophys. J. Lett., 851, L25

\bibitem[{Fishbach \& Holz(2020)}]{Fishbach:2019bbm}
Fishbach, M. \& Holz, D.~E. 2020, Astrophys. J. Lett., 891, L27

\bibitem[{Fishbach {et~al.}(2018)Fishbach, Holz, \& Farr}]{Fishbach:2018edt}
Fishbach, M., Holz, D.~E., \& Farr, W.~M. 2018, Astrophys. J. Lett., 863, L41

\bibitem[{Fragos {et~al.}(2010)Fragos, Tremmel, Rantsiou, \&
  Belczynski}]{Fragos:2010tm}
Fragos, T., Tremmel, M., Rantsiou, E., \& Belczynski, K. 2010, Astrophys. J.
  Lett., 719, L79

\bibitem[{{Franciolini} \& {Pani}(2022)}]{Franciolini:2022iaa}
{Franciolini}, G. \& {Pani}, P. 2022, \prd, 105, 123024

\bibitem[{Fryer \& Kusenko(2006)}]{Fryer:2005sz}
Fryer, C.~L. \& Kusenko, A. 2006, Astrophys. J. Suppl., 163, 335

\bibitem[{Galaudage {et~al.}(2021)Galaudage, Talbot, Nagar, Jain, Thrane, \&
  Mandel}]{Galaudage:2021rkt}
Galaudage, S., Talbot, C., Nagar, T., {et~al.} 2021, Astrophys. J. Lett., 921,
  L15

\bibitem[{Gerosa {et~al.}(2018)Gerosa, Berti, O'Shaughnessy, Belczynski,
  Kesden, Wysocki, \& Gladysz}]{Gerosa:2018wbw}
Gerosa, D., Berti, E., O'Shaughnessy, R., {et~al.} 2018, Phys. Rev. D, 98,
  084036

\bibitem[{Gerosa \& Fishbach(2021)}]{Gerosa:2021mno}
Gerosa, D. \& Fishbach, M. 2021, Nature Astron., 5, 8

\bibitem[{{Giacobbo} \& {Mapelli}(2020)}]{Giacobbo:2019fmo}
{Giacobbo}, N. \& {Mapelli}, M. 2020, \apj, 891, 141

\bibitem[{Golomb \& Talbot(2022)}]{Golomb:2022bon}
Golomb, J. \& Talbot, C. 2022 [\eprint[arXiv]{2210.12287}]

\bibitem[{Gond\'an \& Kocsis(2019)}]{Gondan:2018khr}
Gond\'an, L. \& Kocsis, B. 2019, Astrophys. J., 871, 178

\bibitem[{Hinder {et~al.}(2008)Hinder, Vaishnav, Herrmann, Shoemaker, \&
  Laguna}]{PhysRevD.77.081502}
Hinder, I., Vaishnav, B., Herrmann, F., Shoemaker, D.~M., \& Laguna, P. 2008,
  Phys. Rev. D, 77, 081502

\bibitem[{Hurley {et~al.}(2002)Hurley, Tout, \& Pols}]{Hurley:2002rf}
Hurley, J.~R., Tout, C.~A., \& Pols, O.~R. 2002, Mon. Not. Roy. Astron. Soc.,
  329, 897

\bibitem[{{Hut}(1981)}]{1981A&A....99..126H}
{Hut}, P. 1981, \aap, 99, 126

\bibitem[{{Janka} \& {Mueller}(1994)}]{1994A&A...290..496J}
{Janka}, H.~T. \& {Mueller}, E. 1994, \aap, 290, 496

\bibitem[{Kalogera(2000)}]{Kalogera:1999tq}
Kalogera, V. 2000, Astrophys. J., 541, 319

\bibitem[{{Katz}(1975)}]{1975Natur.253..698K}
{Katz}, J.~I. 1975, \nat, 253, 698

\bibitem[{Mandel {et~al.}(2017)Mandel, Farr, Colonna, Stevenson, Ti\v{n}o, \&
  Veitch}]{Mandel:2016prl}
Mandel, I., Farr, W.~M., Colonna, A., {et~al.} 2017, Mon. Not. Roy. Astron.
  Soc., 465, 3254

\bibitem[{Mandel {et~al.}(2019)Mandel, Farr, \& Gair}]{Mandel:2018mve}
Mandel, I., Farr, W.~M., \& Gair, J.~R. 2019, Mon. Not. Roy. Astron. Soc., 486,
  1086

\bibitem[{Mapelli \& Giacobbo(2018)}]{Mapelli:2018wys}
Mapelli, M. \& Giacobbo, N. 2018, Mon. Not. Roy. Astron. Soc., 479, 4391

\bibitem[{Miller {et~al.}(2020)Miller, Callister, \& Farr}]{Miller:2020zox}
Miller, S., Callister, T.~A., \& Farr, W. 2020, Astrophys. J., 895, 128

\bibitem[{{Mirabel} {et~al.}(2001){Mirabel}, {Dhawan}, {Mignani}, {Rodrigues},
  \& {Guglielmetti}}]{2001Natur.413..139M}
{Mirabel}, I.~F., {Dhawan}, V., {Mignani}, R.~P., {Rodrigues}, I., \&
  {Guglielmetti}, F. 2001, \nat, 413, 139

\bibitem[{{Mirabel} {et~al.}(2002){Mirabel}, {Mignani}, {Rodrigues}, {Combi},
  {Rodr{\'\i}guez}, \& {Guglielmetti}}]{2002A&A...395..595M}
{Mirabel}, I.~F., {Mignani}, R., {Rodrigues}, I., {et~al.} 2002, \aap, 395, 595

\bibitem[{Morscher {et~al.}(2015)Morscher, Pattabiraman, Rodriguez, Rasio, \&
  Umbreit}]{Morscher:2014doa}
Morscher, M., Pattabiraman, B., Rodriguez, C., Rasio, F.~A., \& Umbreit, S.
  2015, Astrophys. J., 800, 9

\bibitem[{Mould \& Gerosa(2022)}]{Mould:2021xst}
Mould, M. \& Gerosa, D. 2022, Phys. Rev. D, 105, 024076

\bibitem[{Mould {et~al.}(2022)Mould, Gerosa, Broekgaarden, \&
  Steinle}]{Mould:2022xeu}
Mould, M., Gerosa, D., Broekgaarden, F.~S., \& Steinle, N. 2022, Arxiv
  preprints [\eprint[arXiv]{2205.12329}]

\bibitem[{{Nelemans} {et~al.}(1999){Nelemans}, {Tauris}, \& {van den
  Heuvel}}]{1999A&A...352L..87N}
{Nelemans}, G., {Tauris}, T.~M., \& {van den Heuvel}, E.~P.~J. 1999, \aap, 352,
  L87

\bibitem[{{Nitz} {et~al.}(2021){Nitz}, {Kumar}, {Wang}, {Kastha}, {Wu},
  {Sch{\"a}fer}, {Dhurkunde}, \& {Capano}}]{Nitz:2021zwj}
{Nitz}, A.~H., {Kumar}, S., {Wang}, Y.-F., {et~al.} 2021, arXiv e-prints,
  arXiv:2112.06878

\bibitem[{{Olsen} {et~al.}(2022){Olsen}, {Venumadhav}, {Mushkin}, {Roulet},
  {Zackay}, \& {Zaldarriaga}}]{Olsen:2022pin}
{Olsen}, S., {Venumadhav}, T., {Mushkin}, J., {et~al.} 2022, \prd, 106, 043009

\bibitem[{{Packet}(1981)}]{1981A&A...102...17P}
{Packet}, W. 1981, \aap, 102, 17

\bibitem[{Peters(1964)}]{PhysRev.136.B1224}
Peters, P.~C. 1964, Phys. Rev., 136, B1224

\bibitem[{Portegies~Zwart \& McMillan(2002)}]{PortegiesZwart:2002iks}
Portegies~Zwart, S.~F. \& McMillan, S. L.~W. 2002, Astrophys. J., 576, 899

\bibitem[{Repetto {et~al.}(2012)Repetto, Davies, \&
  Sigurdsson}]{Repetto:2012bs}
Repetto, S., Davies, M.~B., \& Sigurdsson, S. 2012, Mon. Not. Roy. Astron.
  Soc., 425, 2799

\bibitem[{Rinaldi \& Del~Pozzo(2021)}]{Rinaldi:2021bhm}
Rinaldi, S. \& Del~Pozzo, W. 2021, Mon. Not. Roy. Astron. Soc., 509, 5454

\bibitem[{Rodriguez {et~al.}(2018{\natexlab{a}})Rodriguez, Amaro-Seoane,
  Chatterjee, Kremer, Rasio, Samsing, Ye, \& Zevin}]{Rodriguez:2018pss}
Rodriguez, C.~L., Amaro-Seoane, P., Chatterjee, S., {et~al.}
  2018{\natexlab{a}}, Phys. Rev. D, 98, 123005

\bibitem[{Rodriguez {et~al.}(2018{\natexlab{b}})Rodriguez, Amaro-Seoane,
  Chatterjee, \& Rasio}]{Rodriguez:2017pec}
Rodriguez, C.~L., Amaro-Seoane, P., Chatterjee, S., \& Rasio, F.~A.
  2018{\natexlab{b}}, Phys. Rev. Lett., 120, 151101

\bibitem[{Rodriguez {et~al.}(2015)Rodriguez, Morscher, Pattabiraman,
  Chatterjee, Haster, \& Rasio}]{Rodriguez:2015oxa}
Rodriguez, C.~L., Morscher, M., Pattabiraman, B., {et~al.} 2015, Phys. Rev.
  Lett., 115, 051101, [Erratum: Phys.Rev.Lett. 116, 029901 (2016)]

\bibitem[{Rodriguez {et~al.}(2019)Rodriguez, Zevin, Amaro-Seoane, Chatterjee,
  Kremer, Rasio, \& Ye}]{Rodriguez:2019huv}
Rodriguez, C.~L., Zevin, M., Amaro-Seoane, P., {et~al.} 2019, Phys. Rev. D,
  100, 043027

\bibitem[{Romero-Shaw {et~al.}(2020)}]{Romero-Shaw:2020owr}
Romero-Shaw, I.~M. {et~al.} 2020, Mon. Not. Roy. Astron. Soc., 499, 3295

\bibitem[{Roulet {et~al.}(2021)Roulet, Chia, Olsen, Dai, Venumadhav, Zackay, \&
  Zaldarriaga}]{Roulet:2021hcu}
Roulet, J., Chia, H.~S., Olsen, S., {et~al.} 2021, Phys. Rev. D, 104, 083010

\bibitem[{Roulet \& Zaldarriaga(2019)}]{Roulet:2018jbe}
Roulet, J. \& Zaldarriaga, M. 2019, Mon. Not. Roy. Astron. Soc., 484, 4216

\bibitem[{Safarzadeh {et~al.}(2020)Safarzadeh, Farr, \&
  Ramirez-Ruiz}]{Safarzadeh:2020mlb}
Safarzadeh, M., Farr, W.~M., \& Ramirez-Ruiz, E. 2020, Astrophys. J., 894, 129

\bibitem[{Samsing(2018)}]{Samsing:2017xmd}
Samsing, J. 2018, Phys. Rev. D, 97, 103014

\bibitem[{Speagle(2020)}]{Speagle_2020}
Speagle, J.~S. 2020, MNRAS, 493, 3132

\bibitem[{Stevenson {et~al.}(2017)Stevenson, Vigna-G\'omez, Mandel, Barrett,
  Neijssel, Perkins, \& de~Mink}]{Stevenson:2017tfq}
Stevenson, S., Vigna-G\'omez, A., Mandel, I., {et~al.} 2017, Nature Commun., 8,
  14906

\bibitem[{Talbot {et~al.}(2019)Talbot, Smith, Thrane, \&
  Poole}]{Talbot:2019okv}
Talbot, C., Smith, R., Thrane, E., \& Poole, G.~B. 2019, Phys. Rev. D, 100,
  043030

\bibitem[{Talbot \& Thrane(2017)}]{Talbot:2017yur}
Talbot, C. \& Thrane, E. 2017, Phys. Rev. D, 96, 023012

\bibitem[{Talbot \& Thrane(2018)}]{Talbot:2018cva}
Talbot, C. \& Thrane, E. 2018, Astrophys. J., 856, 173

\bibitem[{Tiwari(2021)}]{Tiwari:2020vym}
Tiwari, V. 2021, Class. Quant. Grav., 38, 155007

\bibitem[{Tong {et~al.}(2022)Tong, Galaudage, \& Thrane}]{Tong:2022iws}
Tong, H., Galaudage, S., \& Thrane, E. 2022, Arxiv preprints
  [\eprint[arXiv]{2209.02206}]

\bibitem[{{Tutukov} \& {Yungelson}(1993)}]{1993MNRAS.260..675T}
{Tutukov}, A.~V. \& {Yungelson}, L.~R. 1993, \mnras, 260, 675

\bibitem[{Vitale {et~al.}(2019)Vitale, Farr, Ng, \& Rodriguez}]{Vitale:2018yhm}
Vitale, S., Farr, W.~M., Ng, K., \& Rodriguez, C.~L. 2019, Astrophys. J. Lett.,
  886, L1

\bibitem[{{Vitale} {et~al.}(2020){Vitale}, {Gerosa}, {Farr}, \&
  {Taylor}}]{Vitale:2020aaz}
{Vitale}, S., {Gerosa}, D., {Farr}, W.~M., \& {Taylor}, S.~R. 2020, arXiv
  e-prints, arXiv:2007.05579

\bibitem[{Vitale {et~al.}(2017)Vitale, Lynch, Sturani, \&
  Graff}]{Vitale:2015tea}
Vitale, S., Lynch, R., Sturani, R., \& Graff, P. 2017, Class. Quant. Grav., 34,
  03LT01

\bibitem[{Wong {et~al.}(2021)Wong, Breivik, Kremer, \&
  Callister}]{Wong:2020ise}
Wong, K. W.~K., Breivik, K., Kremer, K., \& Callister, T. 2021, Phys. Rev. D,
  103, 083021

\bibitem[{Wong {et~al.}(2014)Wong, Valsecchi, Ansari, Fragos, Glebbeek,
  Kalogera, \& McClintock}]{Wong:2013vya}
Wong, T.-W., Valsecchi, F., Ansari, A., {et~al.} 2014, Astrophys. J., 790, 119

\bibitem[{Wysocki {et~al.}(2019)Wysocki, Lange, \&
  O'Shaughnessy}]{Wysocki:2018}
Wysocki, D., Lange, J., \& O'Shaughnessy, R. 2019, Phys. Rev. D, 100, 043012

\bibitem[{Zaldarriaga {et~al.}(2018)Zaldarriaga, Kushnir, \&
  Kollmeier}]{Zaldarriaga:2017qkw}
Zaldarriaga, M., Kushnir, D., \& Kollmeier, J.~A. 2018, Mon. Not. Roy. Astron.
  Soc., 473, 4174

\bibitem[{Zevin {et~al.}(2021)Zevin, Bavera, Berry, Kalogera, Fragos, Marchant,
  Rodriguez, Antonini, Holz, \& Pankow}]{Zevin:2020gbd}
Zevin, M., Bavera, S.~S., Berry, C. P.~L., {et~al.} 2021, Astrophys. J., 910,
  152

\bibitem[{Zevin {et~al.}(2017)Zevin, Pankow, Rodriguez, Sampson, Chase,
  Kalogera, \& Rasio}]{Zevin:2017evb}
Zevin, M., Pankow, C., Rodriguez, C.~L., {et~al.} 2017, Astrophys. J., 846, 82

\end{thebibliography}

\appendix
\section{Hierachical inference}\label{Sec.Method}

 We aim to measure the hyper parameters \vl that control the distribution of single-event parameters \vt (the black hole masses, spins, redshifts, etc.) given the dataset \vd consisting of the 69 \catalog BBHs with false alarm ratio smaller than 1 per year---$\vd \equiv  \{d_i, i=1\ldots 69\}$---reported by the LVK collaboration~\citep{LIGOScientific:2021psn}. 
 
The posterior for \vl can be written as~\citep{Mandel:2018mve,Fishbach:2018edt,Vitale:2020aaz}:
\beq
p(\vl | \vd) \propto {\pi(\vl)}  \prod_{i=1}^{69} \frac{p(d_i |\vl) }{\alpha(\vl)}\,. \label{Eq.HyperPostOfLikeFirst} \nn
\eeq
where we have analytically marginalized over the overall merger rate, which is not relevant for our inference. The function $\alpha(\vl)$ represents the detection efficiency, i.e. the \textit{fraction} of BBHs that are detectable, given the population parameters \vl; $\pi(\vl)$ is the prior for the population hyper parameters, and $p(d_i |\vl) $ is the likelihood of the stretch of data containing the i-th BBH. This allows us to account for selection effects and infer the properties of the underlying, rather than the observed, population.

Using Bayes' theorem and marginalizing over the single-event parameters, the single-event likelihood can be written as 
\begin{equation}
p(d_i |\vl) =\int \ud \vt p(d_i |\vt ) \pi(\vt|\vl) \propto  \int \ud \vt\;\; \frac{  p(\vt | d_i, \maHP)\pi(\vt|\vl) }{\pi(\vt|\maHP)}, \label{eq.Sampling}
\end{equation}
where $p(\vt | d_i, \maHP)$ is the posterior distribution for the binary parameters \vt of the i-th source. The population hyperparameters \vl are typically inferred using a hierarchical process that first involves obtaining posteriors for \vt for each individual event under a non-informative prior, $\pi(\vt|\maHP)$. The hypothesis \maHP represents the settings that were used during this individual-event parameter estimation step. The last term, $\pi(\vt|\vl)$ is the population prior, i.e., our model for how the parameters \vt are distributed in the population, given the hyper parameters.

The integral in Eq.~\ref{eq.Sampling} can be approximated as a discrete sum 
\beq
\int \ud \vt p(d_i |\vt ) \pi(\vt|\vl) \simeq {N_{\rm{samples}}}^{-1} \sum_{k}^{N_{\rm{samples}}}\frac{\pi(\vt^k_i|\vl) }{\pi(\vt^k_i|\maHP)} \nn
\eeq
where the ${N_{\rm{samples}}}$ samples are drawn from the posterior distribution of the i-th event. We use the posterior samples of the 69 BBHs reported in \catalog, as released in \cite{GWTC1Release,GWTC2Release,ligo_scientific_collaboration_and_virgo_2021_5117703,ligo_scientific_collaboration_and_virgo_2021_5546663}. For the sources reported in GWTC-1, we use the samples labelled \texttt{IMRPhenomPv2\_posterior} in the data release; for GWTC-2 we use \texttt{PublicationSamples}; for GWTC-2.1 we use \texttt{PrecessingSpinIMRHM}, and for GWTC-3 we use \texttt{C01:Mixed}. To sample the hyper posterior we use the \textsc{dynesty}~\citep{Speagle_2020} sampler available with the  \texttt{GWPopulation} package~\citep{Talbot:2019okv}.

The detection efficiency $\alpha(\vl)$ can also be calculated through an approximated sum starting from a large collection of simulated BBHs for which the SNR (or another detection statistic) is recorded, as described in \cite{2019RNAAS...3...66F,LIGOScientific:2021psn}. We use the \texttt{endo3\_bbhpop-LIGO-T2100113-v12-\\1238166018-15843600.hdf5} sensitivity file released by the LVK~\citep{ligo_scientific_collaboration_and_virgo_2021_5546676} to calculate $\alpha(\lambda)$, using a false alarm threshold of 1 per year to identify detectable sources, consistently with~\cite{LIGOScientific:2021psn}. 

\section{Reference tilt model}\label{Sec.Reference}

We will be comparing our results against the LVK's model (\lvk) of \citet{LIGOScientific:2021psn}: a mixture between an isotropic component and Gaussian distribution with $\mu=1$ and an unknown standard deviation:
\beq
p(\cos\tau_1,\cos\tau_2 | \sigma, \gau) = \frac{1-\gau}{4} +\gau \prod_j^2{\mathcal{N}(\cos{\tau_j},\mu=1,\sigma)} \label{Eq.LVKModel}
\eeq
The Gaussian component is truncated and normalized in the range $[-1,1]$. The two hyperparameters of \lvk are thus the branching ratio \gau of the Gaussian component and its standard deviation $\sigma$, the same for both black holes. 
We notice that in \cite{Talbot:2017yur} the two normal distributions can assume different values of $\sigma$. However, since in general the spins of the least massive objects are measured with extremely large uncertainty, there are no reasons to expect that imposing the same distribution to both tilts will introduce biases. 

\begin{figure}
\includegraphics[width=0.5\textwidth]{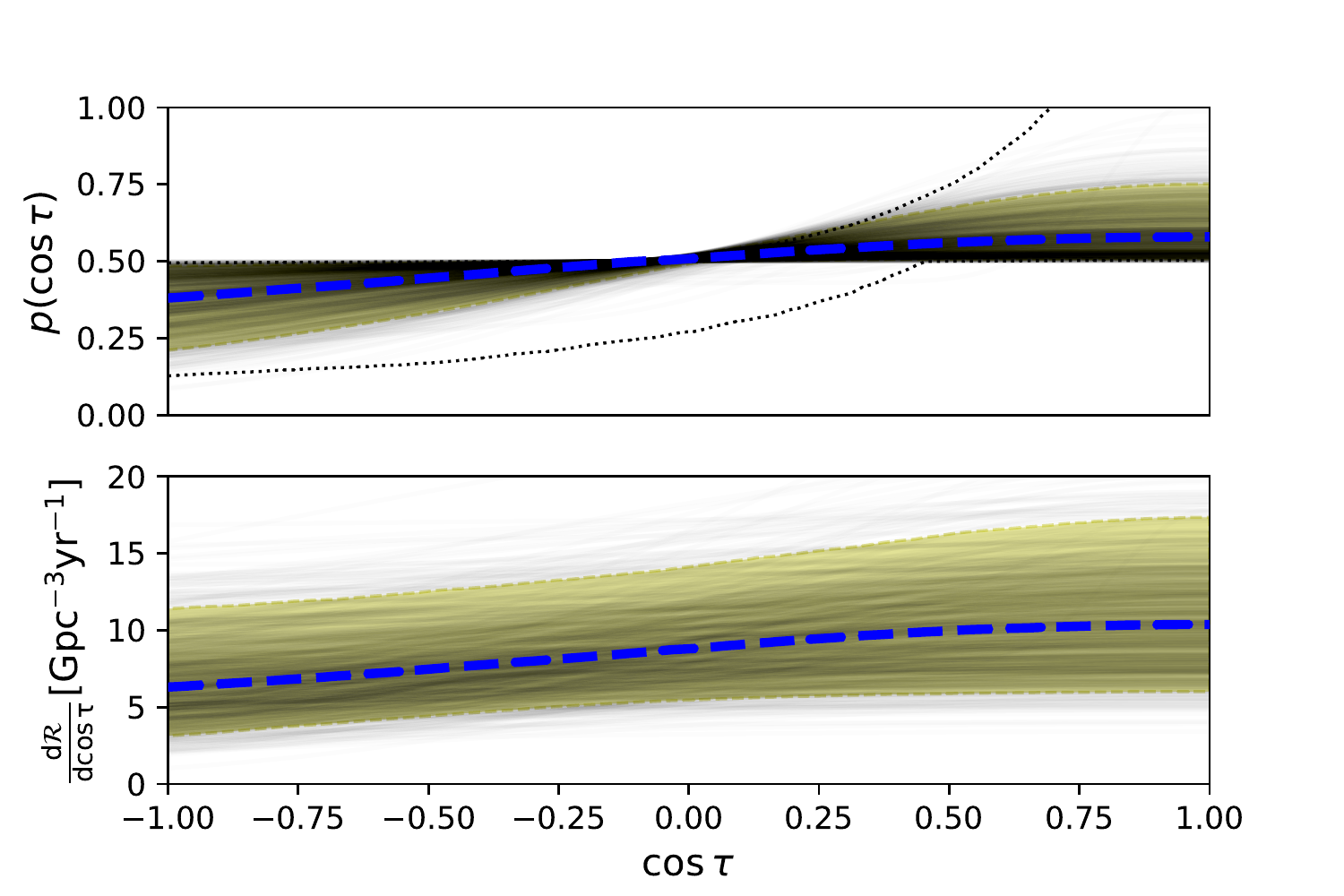}
\caption{(Top) Posterior for \ct obtained using the reference \lvk model. The two thin black dotted lines show 90\% credible interval obtained by drawing the model's hyperparameters from their priors. (Bottom) Differential merger rate per unit \ct for the same model. In both panels, the thin black lines represent individual posterior draws, whereas the colored band shows the 90\% credible interval. The thick dashed line within the band is the median.}
\label{fig.LVK_costau}
\end{figure}

In Fig.~\ref{fig.LVK_costau} we show the resulting inference on the \ct, which---modulo differences in sampling settings---is directly comparable to what is presented by the LVK in \citet{LIGOScientific:2021psn} \add{(their Fig.~15)}. The colored area shows the 90\% credible interval (CI), the thick dashed line is the median, and the dim lines represent individual draws from the posterior. The two dashed lines represent the edges of the 90\% credible interval obtained by sampling the hyperparameters from their priors. It is worth noticing that the \lvk model excludes \textit{a priori} the possibility of an excess of tilts relative to isotropy (i.e. a posterior larger than 0.5) at negative values, as well as a dearth of tilts relative to isotropy for $\ct \gtrsim 0.45$. 
Just as \citet{LIGOScientific:2021psn}, we find that the posterior is not inconsistent with a fully isotropic tilt distribution, while preferring an excess of positive alignment. 

This is shown in the top panel of Fig.~\ref{fig.Ys_comparison}, where the solid blue line is obtained using samples from the hyperparameters' posterior whereas the dashed blue line is obtained by sampling their priors. The fact that there is a hard cutoff at $Y=1$ (the finite bin size causes the curves to extend to values slightly smaller than 1) is  just a symptom of the fact that the \lvk model excludes \textit{a priori} an excess of negative tilts and a dearth at positive tilts, as mentioned above. 

The curve is consistent with $Y=1$, i.e. isotropic posteriors are perfectly consistent with the data, even though it should be appreciated that the model prefers that region \textit{a priori}. The level of consistency can also be assessed with Figure~\ref{fig.iso_gaussian_models_pzeta}, which reports with dashed blue lines the marginalized posterior on the branching ratio of the isotropic component (as opposed to the Gaussian component, to allow direct comparisons with other models). While broad, it favors small values for the fraction of sources in the isotropic component, though fully isotropic distributions ($\iso=1$) are not excluded. The other curves in the figure are discussed in the main body. For all of our models, Tab.~\ref{tab:Evidences} reports the Bayesian evidence, maximum log-likelihood and the number of parameters for the \ct model, as a differential relative to the default LVK model. That table also includes a fully isotropic model (\isoonly, with $p(\cos\tau_1,\cos\tau_2) = 1/4$), which we include as a useful reference. The \isoonly model performs the worst, though not at the point that it can be ruled out with high confidence.

\section{Three-component models}\label{Sec.ThreeCompUncor}

The results presented in Sec.~\ref{sec.Results} show that, depending on the exact model being used, the tilt distribution seems to show either a peak at $+1$, a peak at a smaller positive values of \ct, or a broad plateau for positive \ct.  As this might suggest that two peaks, or features, are present in the data, in this Appendix we consider models that comprise of an isotropic components, plus two other components. To explore the effect of the model on the resulting posterior, we consider different functional forms.

\subsection{\isogbeta model}\label{sec.IsoPlusBetaPlusGaussian}

We use a mixture model with an isotropic component, a Beta distribution component and a Gaussian component: 
\beqa
p(\cos\tau_1,\cos\tau_2 | \gau,\bet,\mu,\sigma,\alpha, \beta) &=& \frac{1-\gau-\bet }{4}+ \nonumber \\ 
+\bet \prod_j^2 \mathcal{B}(\cos \tau_j,\alpha,\beta) &+&\gau \prod_j^2{\mathcal{N}(\cos{\tau_j},\mu,\sigma)}
\label{Eq.IsoPlusGaussianPlusBeta}
\eeqa

In order to reduce degeneracy between the Gaussian and the Beta components, we set the uniform prior for the mean of the Gaussian component to $\mu \sim \mathcal{U}(0.9,5)$; meanwhile we restrict the prior of the Beta parameters to non-singular values, $\alpha,\beta~\sim \mathcal{U}(1,20)$. In practice, this reduces the possibility that the two components can both create peaks in the same region of the \ct domain, which would increase degeneracy and hence make sampling more inefficient. 

\begin{figure}
\includegraphics[width=0.5\textwidth]{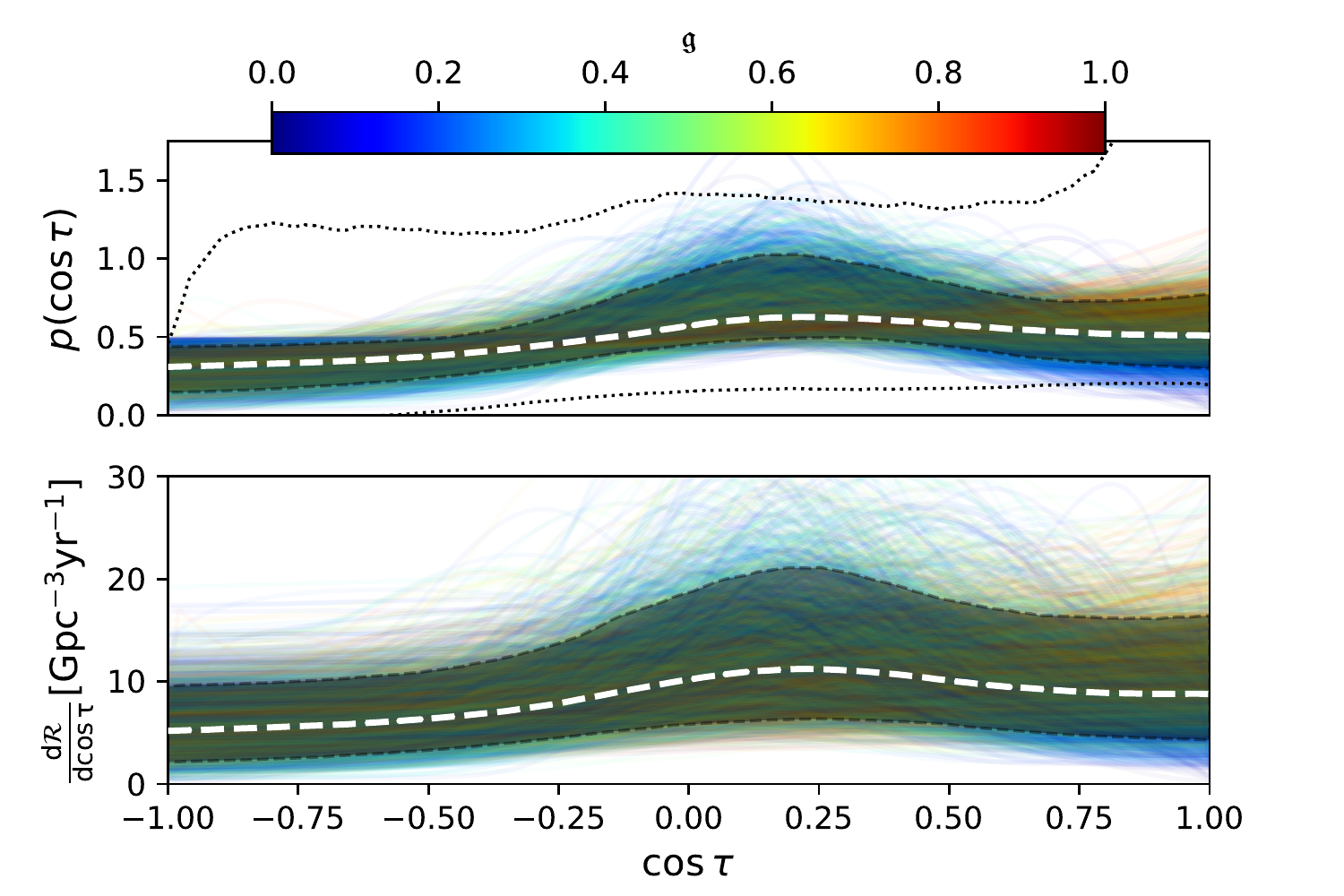}
\caption{Same as Fig.~\ref{fig.IsoPlusTukeyExt_None_costau}, but for the \isogbeta model. Individual posterior draws are colored according to the branching ratio of the Gaussian component, \gau. Note the different scale for the y axis of the bottom panel compared with similar figures for other models.}
\label{fig.IsoPlusBetaPlusGaussian_None_costau}
\end{figure}

\begin{figure}
\includegraphics[width=0.45\textwidth]{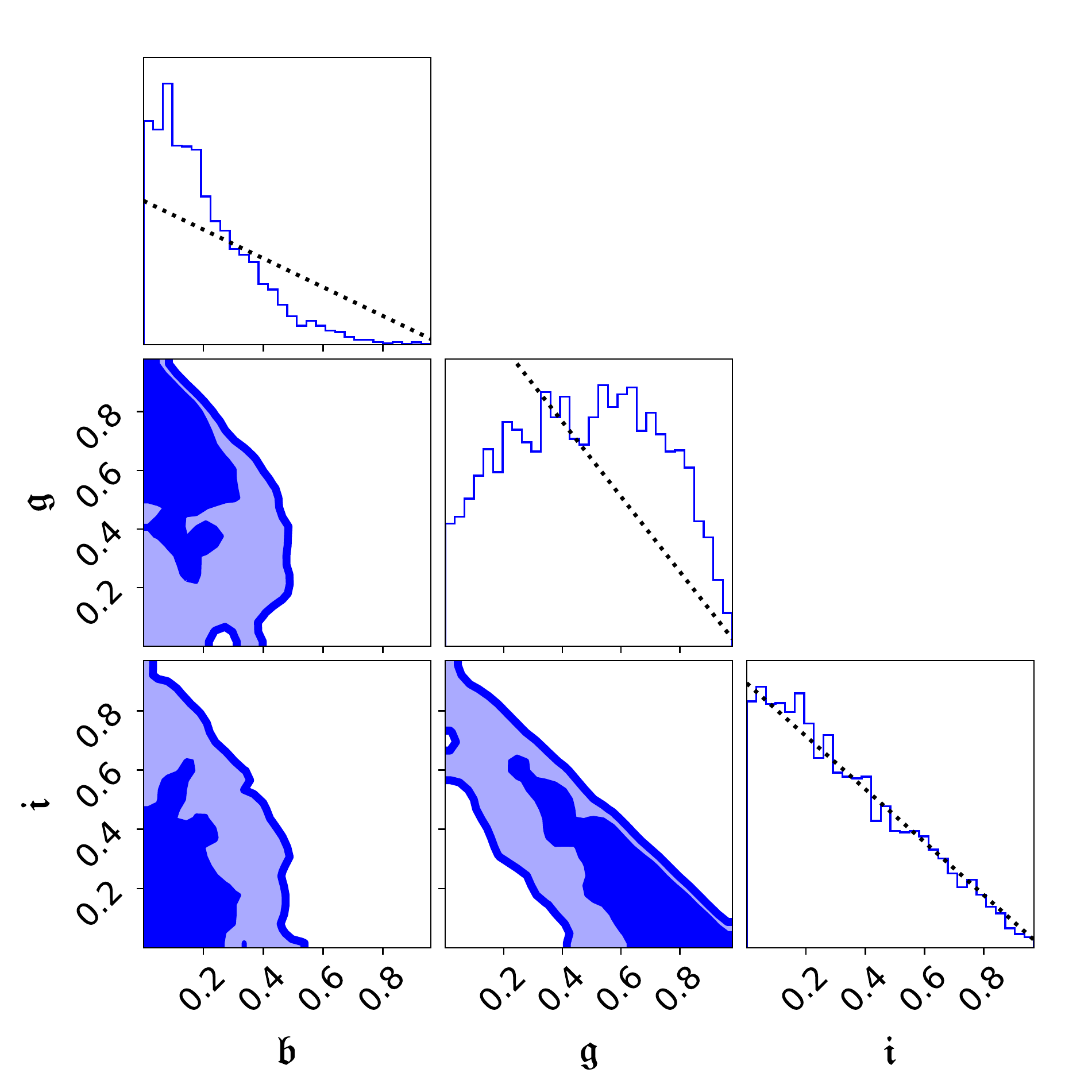}
\caption{Joint and marginal posteriors for the branching ratios of all channels for the \isogbeta model. The thin dashed lines in the diagonal plots are the corresponding priors.}
\label{fig.IsoPlusBetaPlusGaussian_None_branching_ratios}
\end{figure}

The resulting \ct posterior in shown in Fig.~\ref{fig.IsoPlusBetaPlusGaussian_None_costau}. The individual posterior draws are colored according to their value of $\gau$ (we stress that small (large) \gau does \textit{not} necessarily imply large (small) \bet since the isotropic fraction $\iso\equiv 1-\gau-\bet$ needs not be zero). The 90\% CI shows traces of the two features we encountered previously, namely a peak at $+1$ and one at smaller positive values of \ct. As with all of the other models explored in this paper (and, to our knowledge, in the literature) we find that the data excludes an excess of black holes with $\ct \simeq -1 $. The corner plot in Fig.~\ref{fig.IsoPlusBetaPlusGaussian_None_branching_ratios} shows the three branching ratios for this model. The prior for \gau and \bet was uniform in the plane, with the constrain that $\gau+\bet\leq 1$; we show the resulting marginal priors as dotted black lines in the diagonal panels. This model prefers small values of \bet coupled with large values of \gau, as shown in the top-left off-diagonal panel. There is little posterior support for even moderate values of \bet: the 95th percentile for the marginal posterior $p(\bet)$ is $0.50$.
As already visible in Fig.~\ref{fig.IsoPlusBetaPlusGaussian_None_costau}, the Beta component peaks at small positive values of \ct: we find $\mu_\beta = \IsoGaussBetaMu$.

\subsection{\isoggaus model}

Next, we use a mixture model with an isotropic component, and two Gaussian distributions. Here too, to avoid perfect degeneracy, we restrict somewhat the allowed range of the Gaussian means. 
The Gaussian on the right (index ``R'') has a mean that can only vary in the range $[0.9,5]$.
The prior for the Gaussian one on the left (index ``L'') spans the range $[-1,1]$, however, we set to 0 the likelihood for samples for which $\mu_L \notin [A,B]$, where $A$ and $B$ are hyperparameters of the model.
In practice, this implies that the Gaussian on the left is truncated (and hence normalized) in the range $[A,B]$ \textit{and} has a mean in the same range, for each sample. Mathematically:
\beqa
p(\cos\tau_1,\cos\tau_2 | \gau_L,\gau_R,\mu_L,\sigma_L,\mu_R,\sigma_R,A,B) = \frac{1-\gau_L-\gau_R }{4} && \nonumber \\ 
+\gau_L \prod_j^2 {\mathcal{N}_{[A,B]}(\cos{\tau_j},\mu_L,\sigma_L)} +\gau_R \prod_j^2{\mathcal{N}(\cos{\tau_j},\mu_R,\sigma_R)}.&&
\label{Eq.IsoPlusGaussianPlusGaussian}
\eeqa

\begin{figure}
\includegraphics[width=0.5\textwidth]{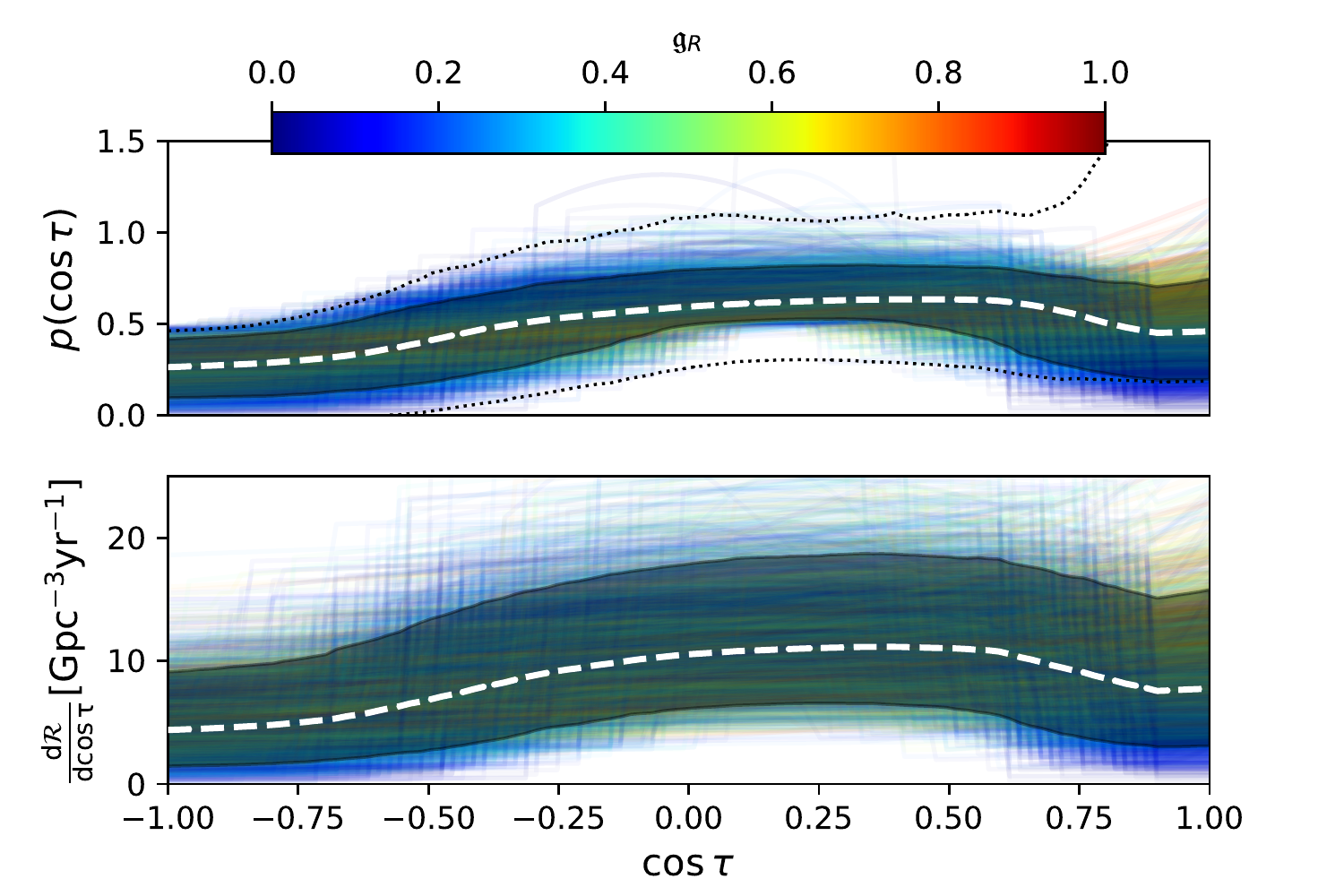}
\caption{Same as Fig.~\ref{fig.IsoPlusTukeyExt_None_costau}, but for the \isoggaus model, when the left Gaussian component is truncated in the range $[A,B]$, with $A$ and $B$ model's hyperparameters. Individual posterior draws are colored according to the branching ratio of the rightmost Gaussian component. Note the different scale for the y axis of the bottom panel compared with similar figures for other models.}
\label{fig.IsoPlusTwoGaussiansMarg_None_costau}
\end{figure}

We explicitly add hyperparameters for the domain of the left Gaussian in order to verify if the data prefers solutions that do not add posterior support to the anti-aligned ($\ct \gtrsim -1$) region \citep[cfr.][]{Callister:2022qwb}.
The resulting posterior for \ct is shown in Fig.~\ref{fig.IsoPlusTwoGaussiansMarg_None_costau}, where the individual posterior draws are colored according to the branching ratio of the right Gaussian -- $\gau_R$. The 90\% CI shows again two features: a rather broad peak for small positive value of \ct and a second peak at $+1$. 

\begin{figure}
\includegraphics[width=0.5\textwidth]{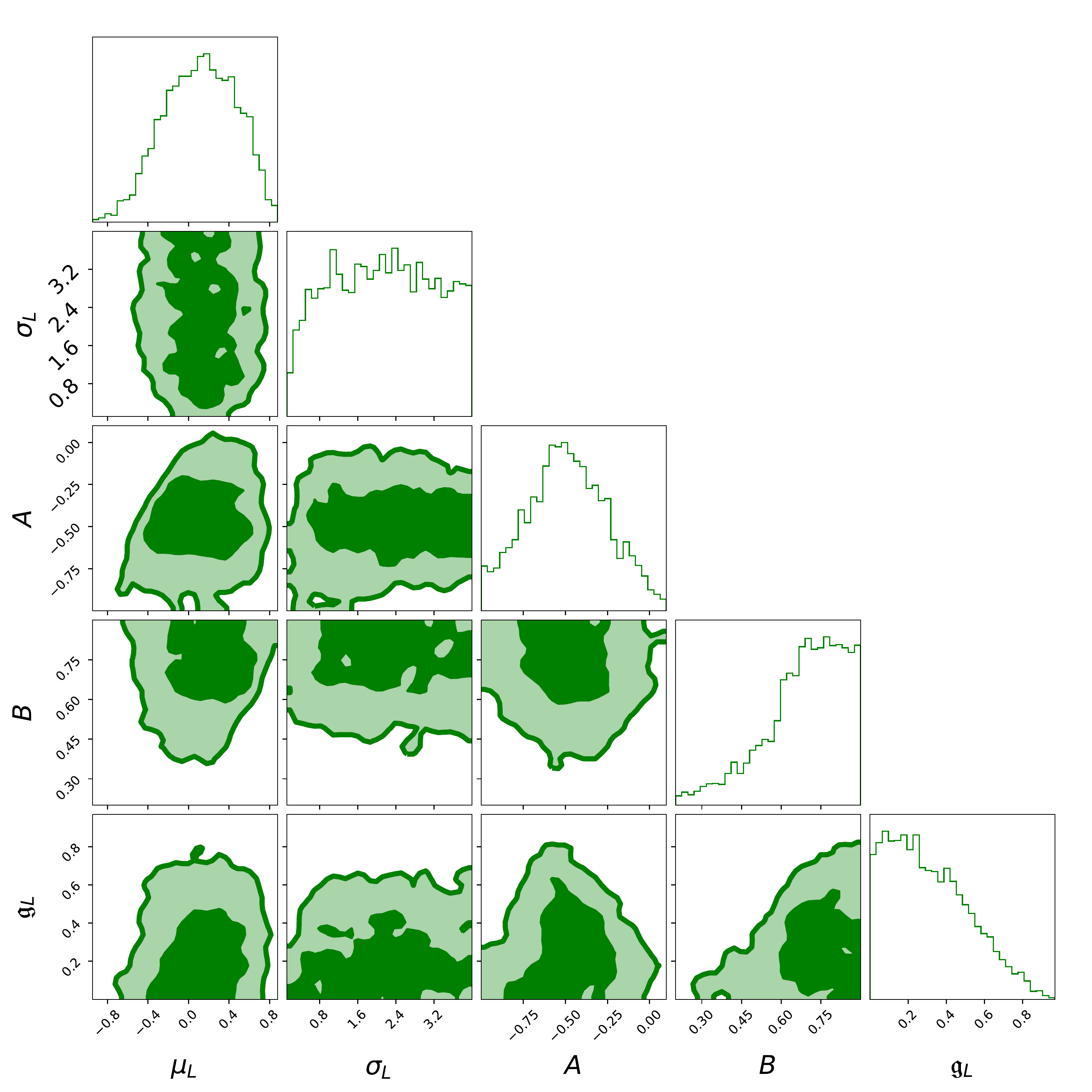}
\caption{Joint and marginal posteriors for the hyperparameters and branching ratio of associated with the left Gaussian of the \isoggaus model.}
\label{fig.IsoPlusTwoGaussiansMarg_None_LeftGaussian}
\end{figure}

We note that the data is informative for the parameters $A$ and $B$, Fig.~\ref{fig.IsoPlusTwoGaussiansMarg_None_LeftGaussian}. While their priors are uniform, the posteriors for both $A$ and $B$ show clear peaks. For $A$ we measure $A= \IsoTwoGaussMargA$, which notably excludes $-1$ at 90\% credibility. Meanwhile, the posterior for $B$ rails against $+1$. The standard deviation for the left Gaussian, $\sigma_L$ is large (the 5th percentile of $p(\sigma_L |d)$ is $\IsoTwoGaussMargSigmaLPercFive$) which implies that even though functionally speaking this component of our model is a Gaussian, the data seems to prefer very wide Gaussians, resembling pieces of segments. The mean of the left Gaussian prefers small positive values, $\mu_L = \IsoTwoGaussMargMuL$, and shows no obvious correlation with $\sigma_L$. %

\begin{figure}
\includegraphics[width=0.45\textwidth]{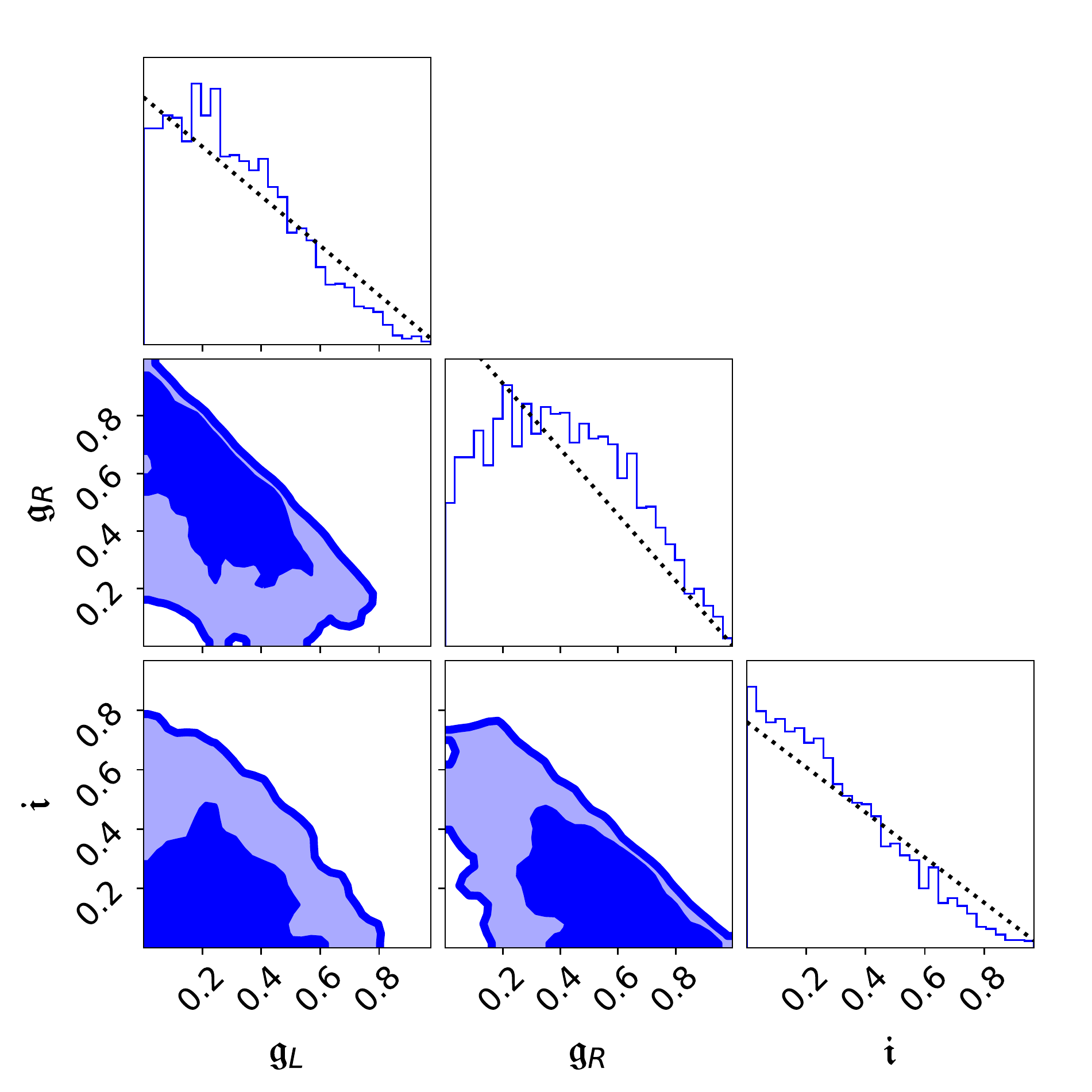}
\caption{Same as Fig.~\ref{fig.IsoPlusBetaPlusGaussian_None_branching_ratios} but for the \isoggaus model.}
\label{fig.IsoPlusTwoGaussiansMarg_None_branching_ratios}
\end{figure}

Figure~\ref{fig.IsoPlusTwoGaussiansMarg_None_branching_ratios} shows the branching ratios for the two Gaussian and the isotropic component $\iso\equiv 1-\gau_R - \gau_L$. together with the corresponding priors (dashed lines). The measurement is not precise, and only small departures from the priors are apparent. In particular, for both $\gau_R$ and $\gau_L$ the posteriors yield a wide peak at $\sim 0.5$, whereas \iso peaks at $0$ more than the prior. 

\begin{figure}
\includegraphics[width=0.5\textwidth]{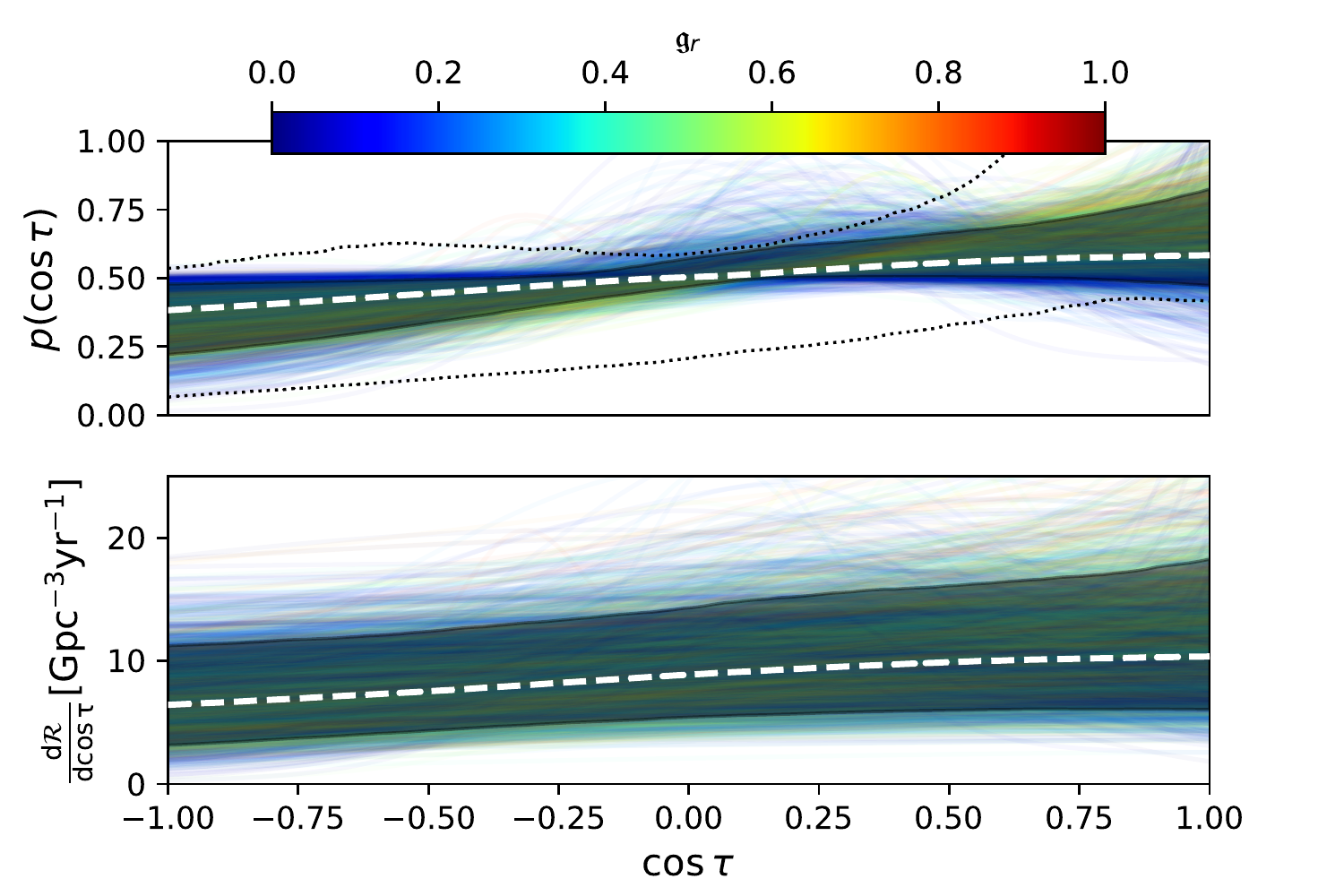}
\caption{Same as Fig.~\ref{fig.IsoPlusTwoGaussiansMarg_None_costau}, but without truncating the left Gaussian (i.e. with $-A=B=1$). Individual posterior draws are colored according to the branching ratio of the rightmost Gaussian component. Note the different scale for the y axis of the bottom panel compared with similar figures for other models.}
\label{fig.IsoPlusTwoGaussiansNoMarg_None_costau}
\end{figure}

We stress that the posterior on \ct we obtained for this model is heavily impacted by the fact that the left Gaussian is truncated in a range, whose position is measured from the data. If instead we set $-A=B=1$, i.e. we extend (and normalize) the left Gaussian to the full \ct range, we obtain a radically different posterior, Fig.~\ref{fig.IsoPlusTwoGaussiansNoMarg_None_costau}. The branching ratio for left Gaussian component in this case doesn't show significant differences relative to the prior. While some of the posterior draws show prominent peaks for small positive values of \ct, those are not frequent enough to create a visible peak in the  90\% CI band, as was instead the case in Fig.~\ref{fig.IsoPlusTwoGaussiansMarg_None_costau}.

Given that the only difference between the models behind Fig.~\ref{fig.IsoPlusTwoGaussiansNoMarg_None_costau} and Fig.~\ref{fig.IsoPlusTwoGaussiansMarg_None_costau} is the truncation of the left Gaussian's domain, it is tempting to think that the tails of the left Gaussian---if free to extend all the way to $\ct=-1$---would give too much posterior weight in that region, which is not supported by the data. This explanation is also consistent with the fact that our model of Sec.~\ref{sec.IsoPlusBetaPlusGaussian} does find the peak, since a Beta distribution can produce tails which are less wide than a Gaussian.

\subsection{\isogtuk model}

To end our exploration of 3-component models, we modify the model of the previous section and replace the left Gaussian with a distribution based on the Tukey window function. Mathematically:
\beqa
p(\cos\tau_1,\cos\tau_2 | \tuk,\gau,T_{x0},T_k,T_r,\mu,\sigma) = \frac{1-\tuk-\gau }{4} && \nonumber \\ 
+\tuk \prod_j^2 {\mathcal{T}(\cos{\tau_j},T_{x0},T_k,T_r)} +\gau \prod_j^2{\mathcal{N}(\cos{\tau_j},\mu,\sigma)}&&
\label{Eq.IsoPlusGaussianPlusTukey}
\eeqa
The priors for all of the hyper-parameters are uniform, with the exception of the branching ratios \tuk and $\gau$, which are jointly uniform in the triangle $\tuk + \gau \leq 1$.

\begin{figure}
\includegraphics[width=0.45\textwidth]{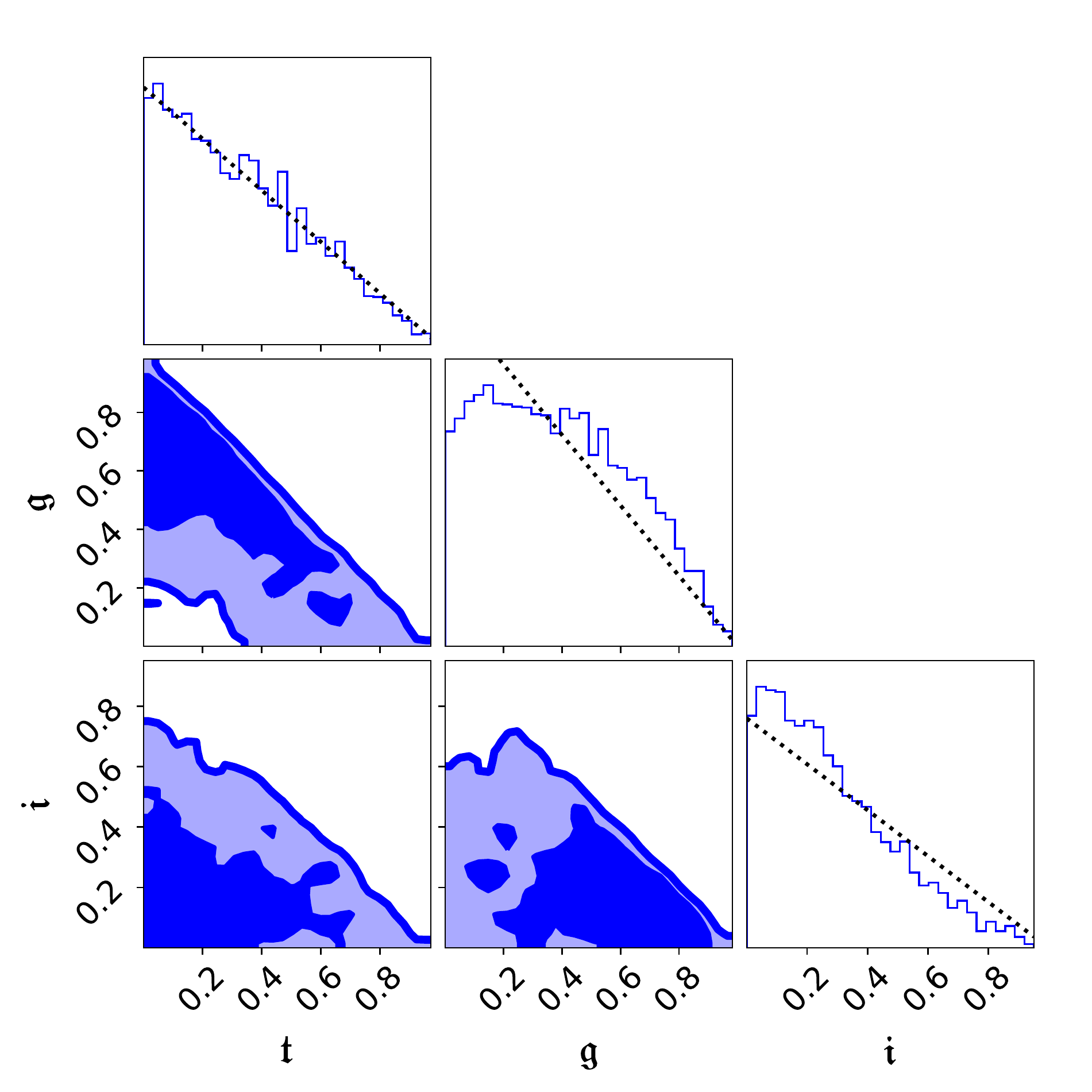}
\caption{Same as Fig.~\ref{fig.IsoPlusBetaPlusGaussian_None_branching_ratios} but for the \isogtuk model.}
\label{fig.IsoPlusTukeyExtPlusGaussian_None_branching_ratios}
\end{figure}

\begin{figure}
\includegraphics[width=0.45\textwidth]{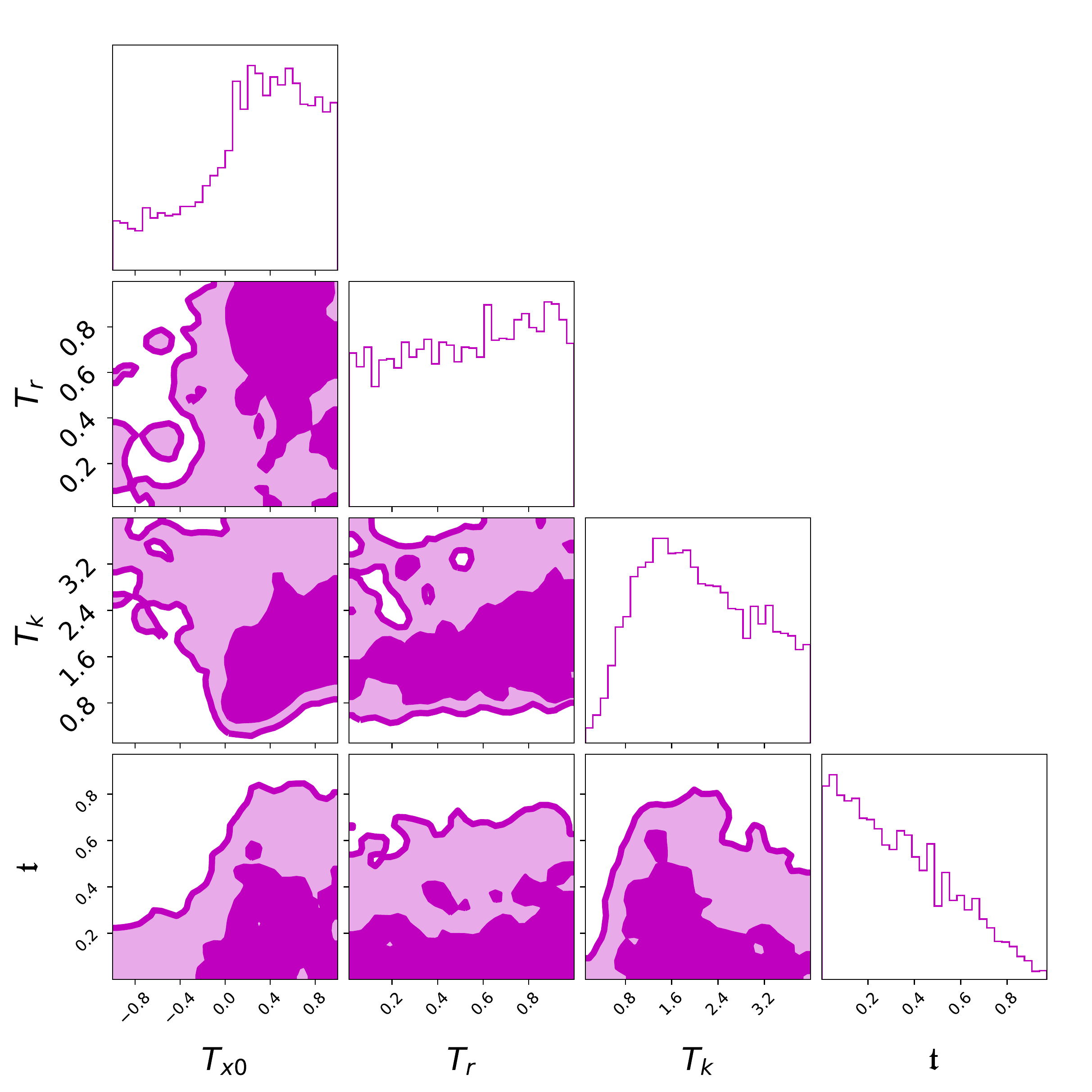}
\caption{Joint and marginal posteriors for the hyperparameters and branching ratio of the Tukey component  of the \isogtuk model.}
\label{fig.IsoPlusTukeyExtPlusGaussian_None_TukeyPars}
\end{figure}

Figure~\ref{fig.IsoPlusTukeyExtPlusGaussian_None_branching_ratios} shows the posteriors for the branching ratios, including that of the isotropic component $\iso\equiv 1 - \tuk - \gau$. As for the \isoggaus model, the branching ratios are not measured with precision. The data is prefers smaller values of \tuk and \iso and moderate values of \gau. 
Comparing Fig.~\ref{fig.IsoPlusTukeyExtPlusGaussian_None_TukeyPars} with the corresponding plot for the \isotuk run---Fig.~\ref{fig.IsoPlusTukeyExt_None_TukeyPars}---we find qualitatively consistent results. In particular, $T_{x0}$ has mostly support at positive values, except when $T_k$ can take large values or \tuk is small. 
Using the full posterior, we find $T_{x0} =\IsoGaussTukeyMu$, while restricting to samples with $T_k\leq2$ ($T_k\leq1$) yields $T_{x0} = \IsoGaussTukeyMuTkLTwo$ ($T_{x0} = \IsoGaussTukeyMuTkLOne$) consistent with the simpler 2-component model.

Similarly, we find that the posterior for \ct mainly differs from that of Fig.~\ref{fig.IsoPlusTukeyExt_None_costau} because of some additional -- but not large -- support at $\ct=1$, due to the contribution of the Gaussian component. 

\begin{figure}
\includegraphics[width=0.5\textwidth]{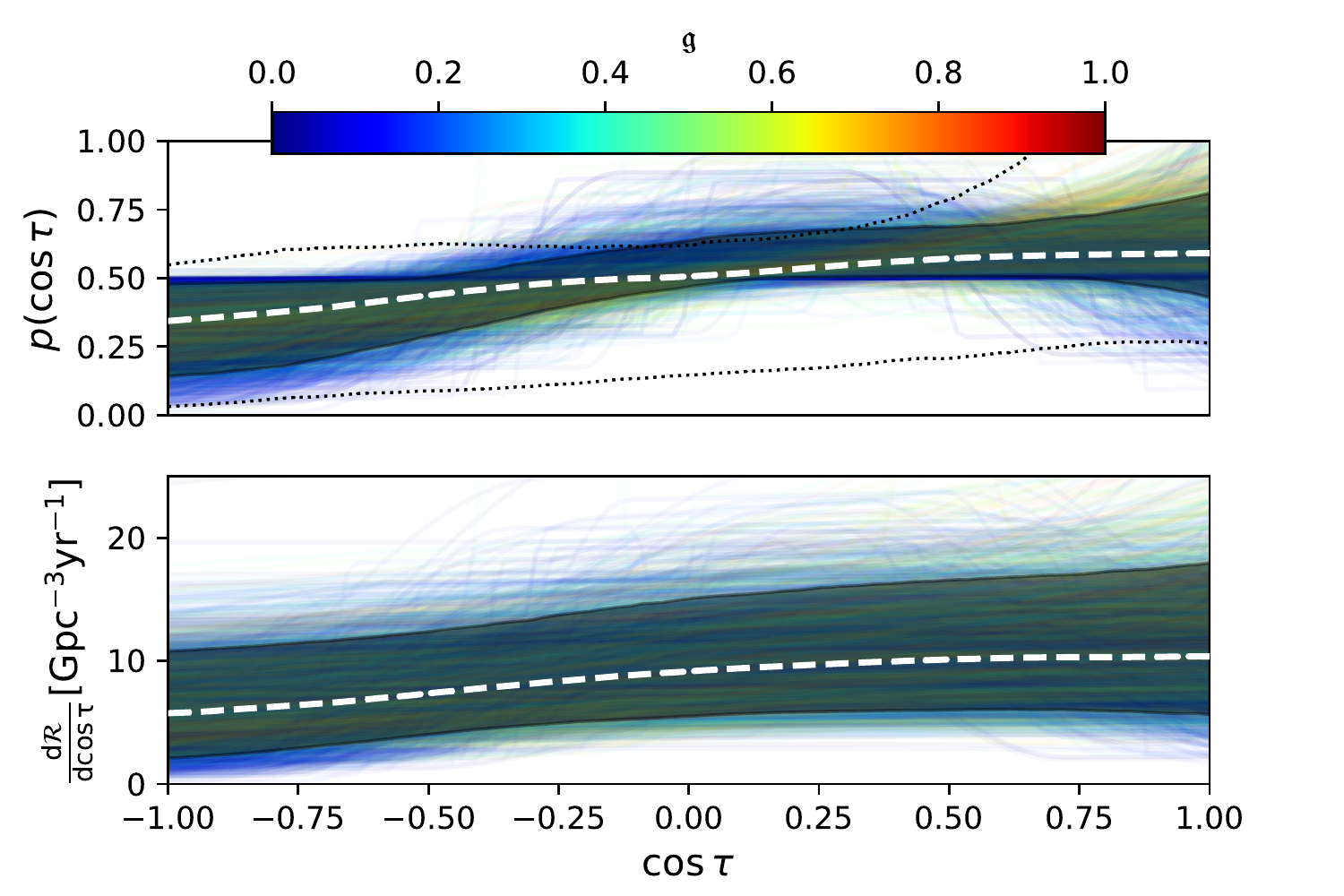}
\caption{Same as Fig.~\ref{fig.IsoPlusTukeyExt_None_costau}, but for the \isogtuk model. Individual posterior draws are colored according to the branching ratio of the Gaussian component. Note the different scale for the y axis of the bottom panel compared with similar figures for other models.}
\label{fig.IsoPlusTukeyExtPlusGaussian_None_costau}
\end{figure}

\section{Correlated mixture models}~\label{Sec.Correlated}

In this Appendix we revisit some of the models presented in the main body of the paper, and we extend them to allow for the possibility that the hyperparameters governing the population-level \ct distribution are correlated with some of the astrophysical binary parameters. Previous works have considered correlations between the effective aligned spin, \chieff, and the BBH masses and redshifts~\citep{Safarzadeh:2020mlb, Callister:2021fpo, Franciolini:2022iaa, Biscoveanu:2022qac,Adamcewicz:2022hce}, but not a direct correlation between the tilts and these other intrinsic parameters. Given that the number of BBH sources is still relatively small, we only consider a subset of 2-component models, in order to keep the number of hyperparameters limited. For some of the correlated models, the distributions for the two tilt angles are not assumed to be identical (this happens when each tilt is allowed to be correlated with the corresponding component mass or spin magnitude): we will only report the distribution for the tilt of primary (i.e. most massive) black hole, as it is usually best measured.

\subsection{\isocgaus model}

We first allow for the possibility that the mean and standard deviation of the Gaussian component might be correlated with other astrophysical parameters -- $\kappa$, described below -- since those should be related to the details of the supernovae explosions that would have tilted the orbit~\citep[e.g.,][]{Gerosa:2018wbw}. We minimally modify the \isogaus model to allow the mean and standard deviation to linearly vary with the parameter that is correlated to the spin tilt. This introduces another set of hyper parameters, which control the linear part of the mean and standard deviation:
\beqa
p(\cos\tau_1,\cos\tau_2 | \mu_a,\mu_b ,\sigma_a,\sigma_b,\gau, \kappa_1,\kappa_2) &=& \frac{1-\gau}{4} \nn \\
+\gau \prod_j^2{\mathcal{N}(\cos{\tau_j},\mu(\kappa_j),\sigma(\kappa_j))}, &&
\label{Eq.IsoPlusCorrelatedGaussian}
\eeqa
where $\mu(\kappa) = \mu_a + \frac{\kappa}{N} \mu_b$ and $\sigma(\kappa) = \sigma_a + \frac{\kappa}{N} \sigma_b$.

Notice that even though we could also have allowed for correlations in the branching ratio \gau, we decided not to, as that parameter is already very poorly measured, cfr. Fig~\ref{fig.iso_gaussian_models_pzeta}. Similarly, and following \cite{Callister:2021fpo,Biscoveanu:2020are} we only consider linear correlations. As more sources are detected, both of these assumptions might be trivially relaxed. The constant $N$ is chosen to guarantee that the coefficient of $\mu_b, \sigma_b$ is always smaller than 1. We explore the following possible correlations:

\begin{itemize}
\item \textbf{Component masses, $\kappa_1=m_1,\kappa_2= m_2$ $N=100~\msun$}  
\item \textbf{Component spins, $\kappa_1= \chi_1,\kappa_2=\chi_2, N=1$}  
\item \textbf{Mass ratio, $\kappa_1=\kappa_2= q, N=1$}  
\item \textbf{Total mass, $\kappa_1=\kappa_2= m_{\rm{tot}}, N=200~\msun$}  
\end{itemize}

\begin{figure}
\includegraphics[width=0.5\textwidth]{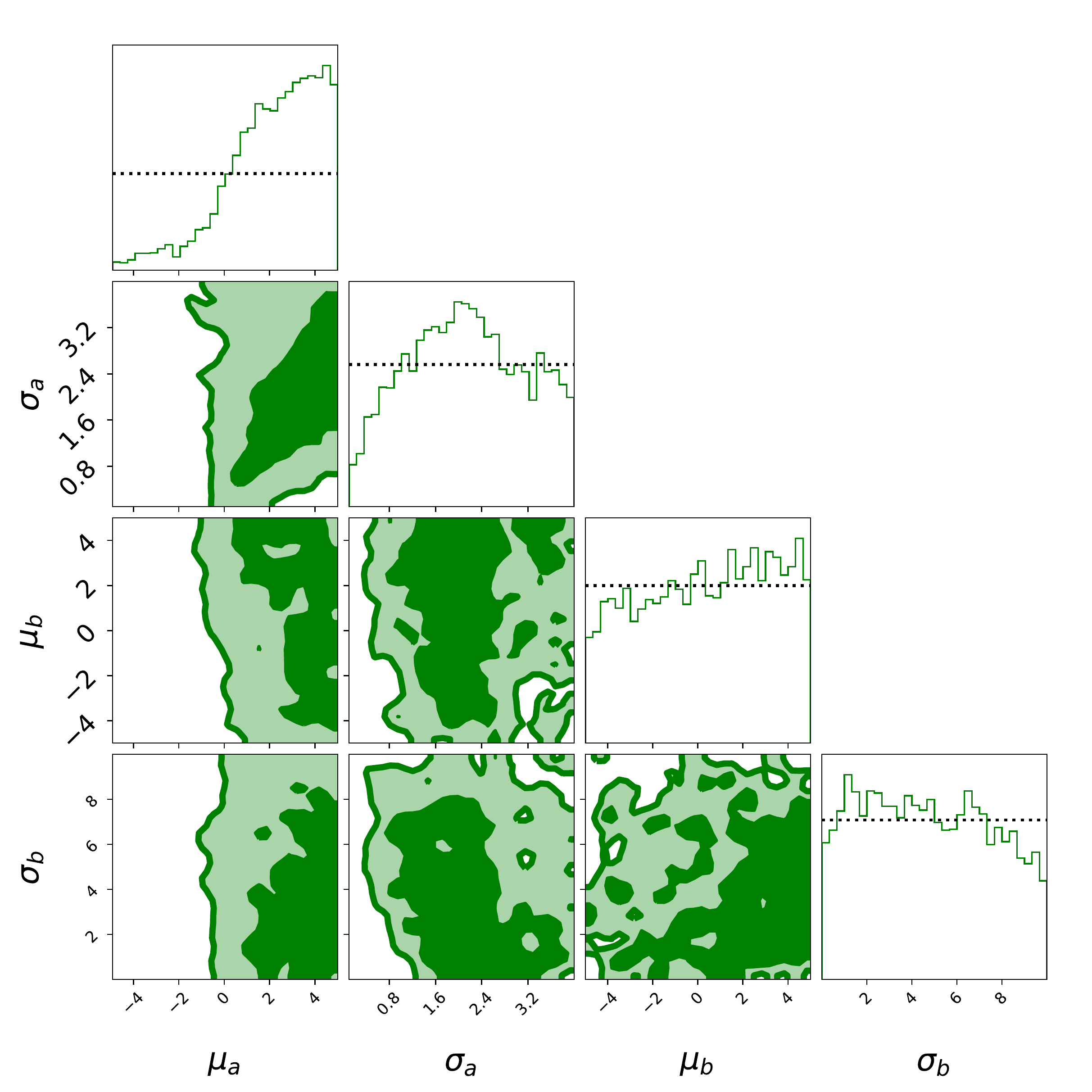}
\caption{Joint and marginal posteriors for the Gaussian parameters obtained in the analysis where they are correlated with the component masses. Dashed lines in the diagonal panels represent the priors. While $\mu_a$ and $\sigma_a$ resemble the corresponding posterior in the \isogaus model, Fig.~\ref{fig.IsoMovingGaussian_None_Combined_sns_mu_sigma}, the terms that enact the correlation, $\mu_b$ and $\sigma_b$ are nearly unconstrained.}
\label{fig.IsoPlusCorrelatedGaussian_m1_GaussianPars}
\end{figure}

\begin{figure}[h]
\includegraphics[width=0.45\textwidth]{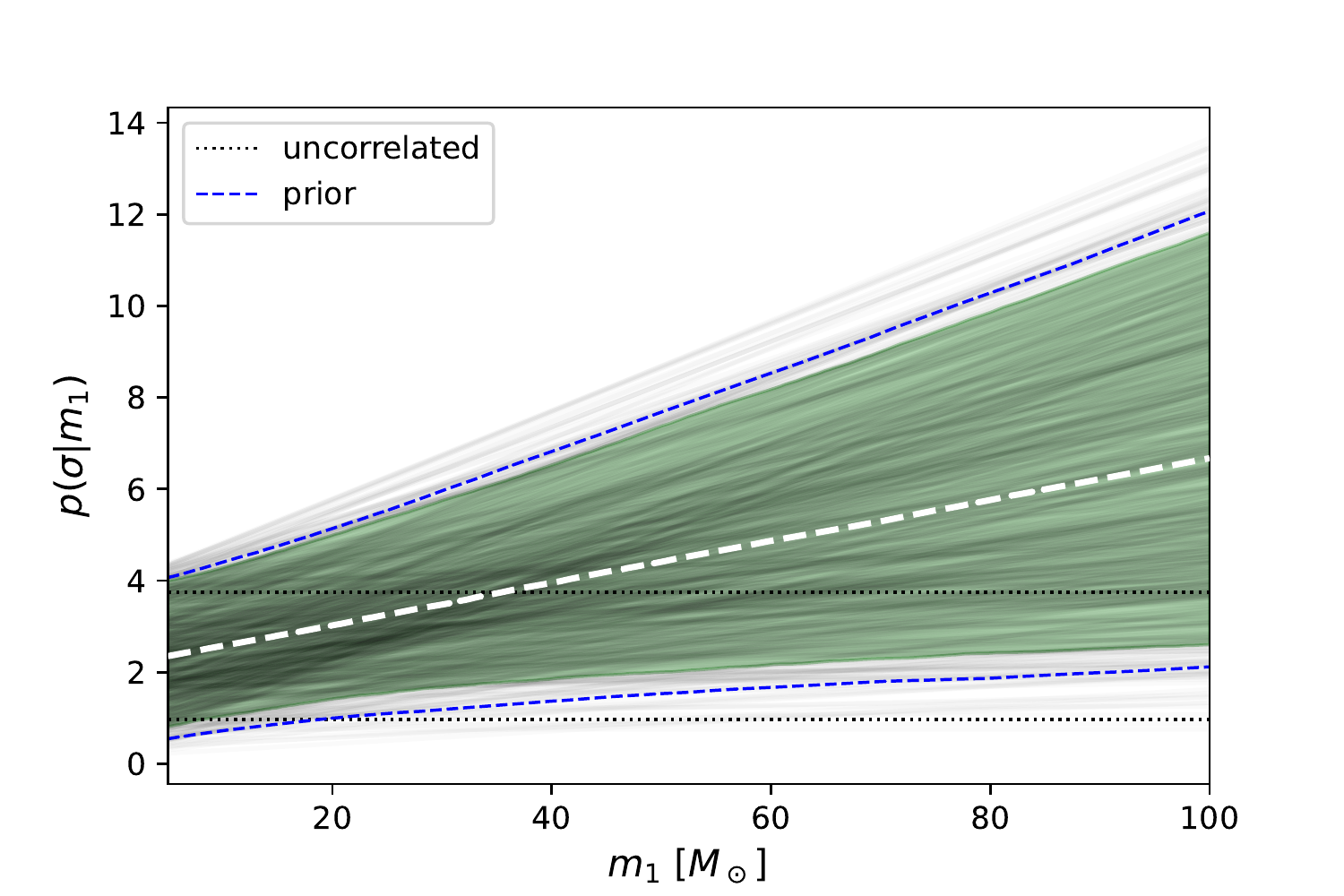}
\caption{Posterior of the standard deviation of the Gaussian component for the \isocgaus model, when we allow for correlation with the component masses, as a function of the primary mass $m_1$. The thin black lines are individual posterior draws, the colored band is the 90\% CI and the thick dashed line is the median. The two horizontal think dotted lines enclose the 90\% CI for the  \isogaus model, which does not allow for correlations. Finally, the two thin blue dashed line enclose the 90\% CI obtained sampling the prior. }
\label{fig.iso_correlated_truncnorm_m1_sigma_of_m1}
\end{figure}

We find that we cannot constrain in any significant way the parameters that enact the correlations (i.e. $\mu_b$ and $\sigma_b$), for which we recover posteriors which highly resemble the corresponding priors. 
This is shown in Fig.~\ref{fig.IsoPlusCorrelatedGaussian_m1_GaussianPars}, where we report the parameters of the Gaussian component for the analysis where they are correlated with the component masses. The dashed horizontal lines in the diagonal panels represent the corresponding priors. Especially for the standard deviation term $\sigma_b$, no information is gained relative to the prior. 
Because we restrict the prior on $\sigma_{b}$ to non-negative values (see Tab.~\ref{tab:priors}) to ensure that the width of the Gaussian does not become negative for any values of $\kappa$, this implies that the overall standard deviation for the Gaussian component can only increase with the mass, Fig.~\ref{fig.iso_correlated_truncnorm_m1_sigma_of_m1}.
However, as made clear by comparing the 90\% CI band with the extent of the 90\% CI obtained with prior draws, the increase of $\sigma$ is entirely prior-driven. 

\subsection{\isocbeta model}

Finally, we augment the \isobeta model of Sec.~\ref{sec.IsoPlusBeta} to allow for correlations in the parameters that control the Beta component:
\beqa
p(\cos\tau_1,\cos\tau_2 | \alpha_a, \alpha_b,\beta_a,\beta_b,\kappa_1,\kappa_2,\bet) &=&\frac{1-\bet}{4} \nn\\
+\bet \prod_j^2 \mathcal{B}(\cos \tau_j,\alpha(\kappa_j),\beta(\kappa_j)),
\label{Eq.IsoPlusCorrelatedBeta}
\eeqa
with $ \alpha(\kappa) = \alpha _a+ \frac{\kappa}{N} \alpha_b$ and $\beta(\kappa) = \beta_a + \frac{\kappa}{N}  \beta_b$.

\begin{figure}
\includegraphics[width=0.45\textwidth]{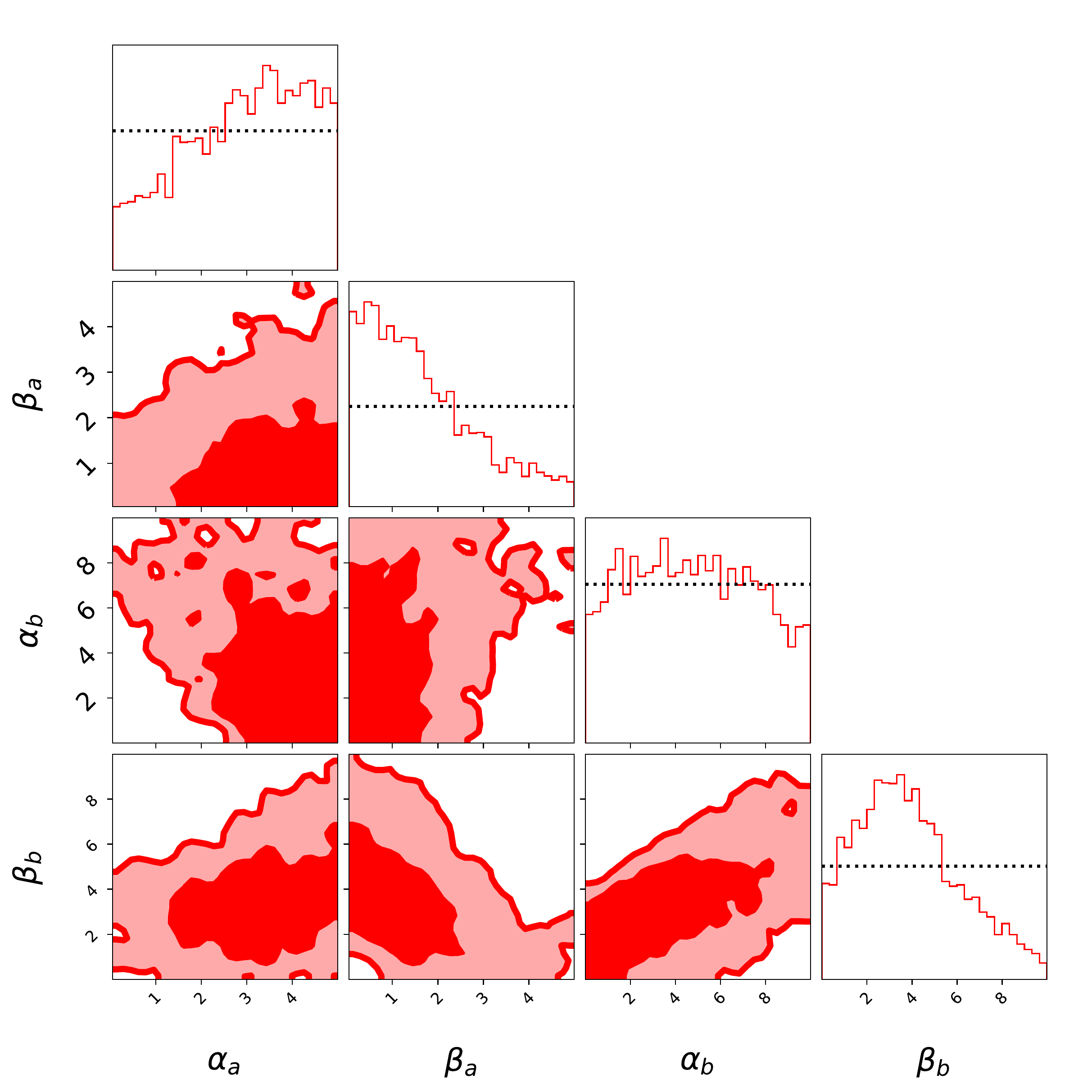}
\caption{Joint and marginal posteriors for the Beta parameters obtained in the analysis where they are correlated with the mass ratio. Dashed lines in the diagonal panels represent the priors.}
\label{fig.IsoPlusCorrelatedBeta_q_BetaPars}
\end{figure}

\begin{figure}
\includegraphics[width=0.45\textwidth]{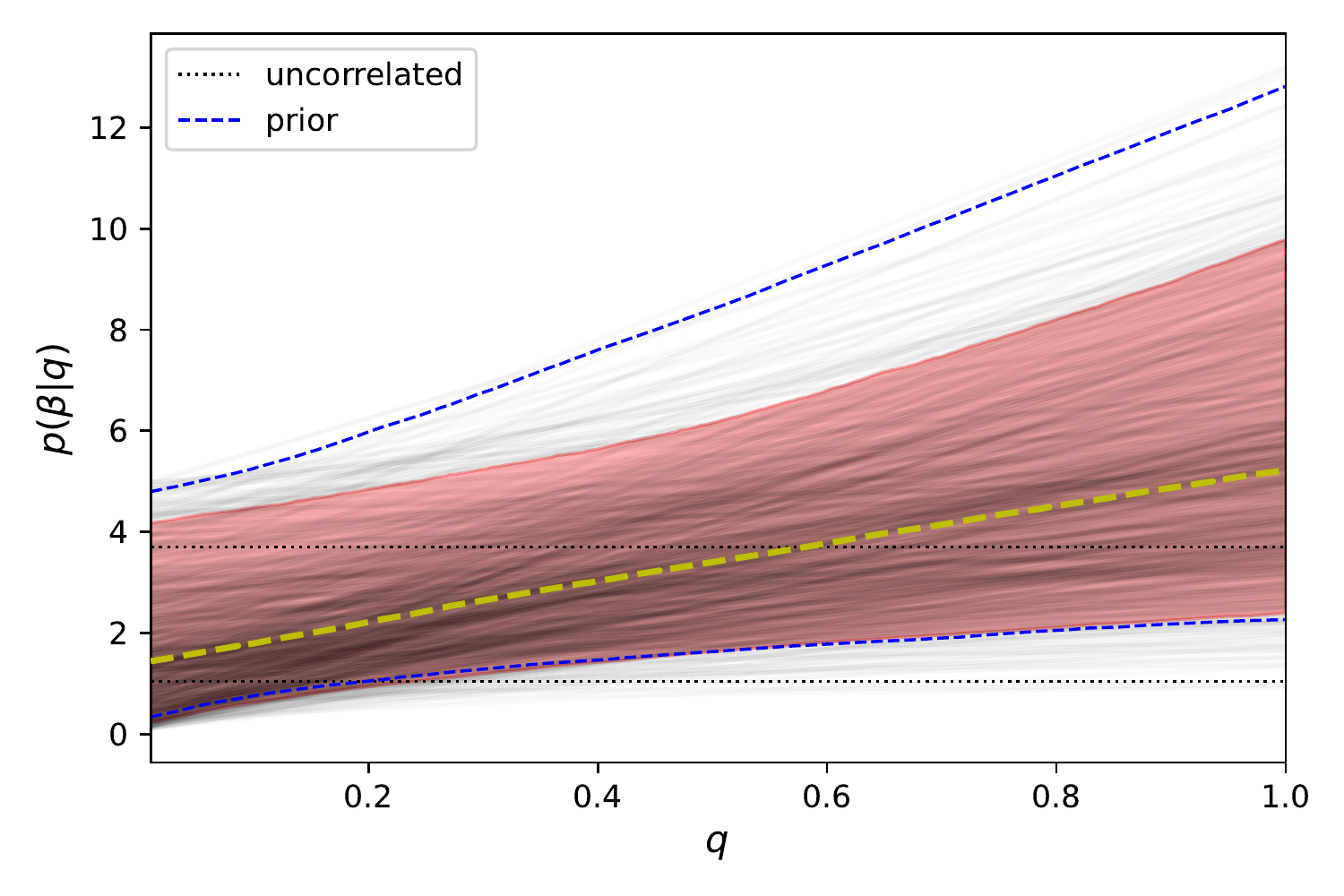}
\caption{Same as ~\ref{fig.iso_correlated_truncnorm_m1_sigma_of_m1} but for the $\beta$ parameter of the \isocbeta model, when correlated with the mass ratio $q$.}
\label{fig.iso_correlated_beta_q_beta_of_q}
\end{figure}

\begin{figure}
\includegraphics[width=0.45\textwidth]{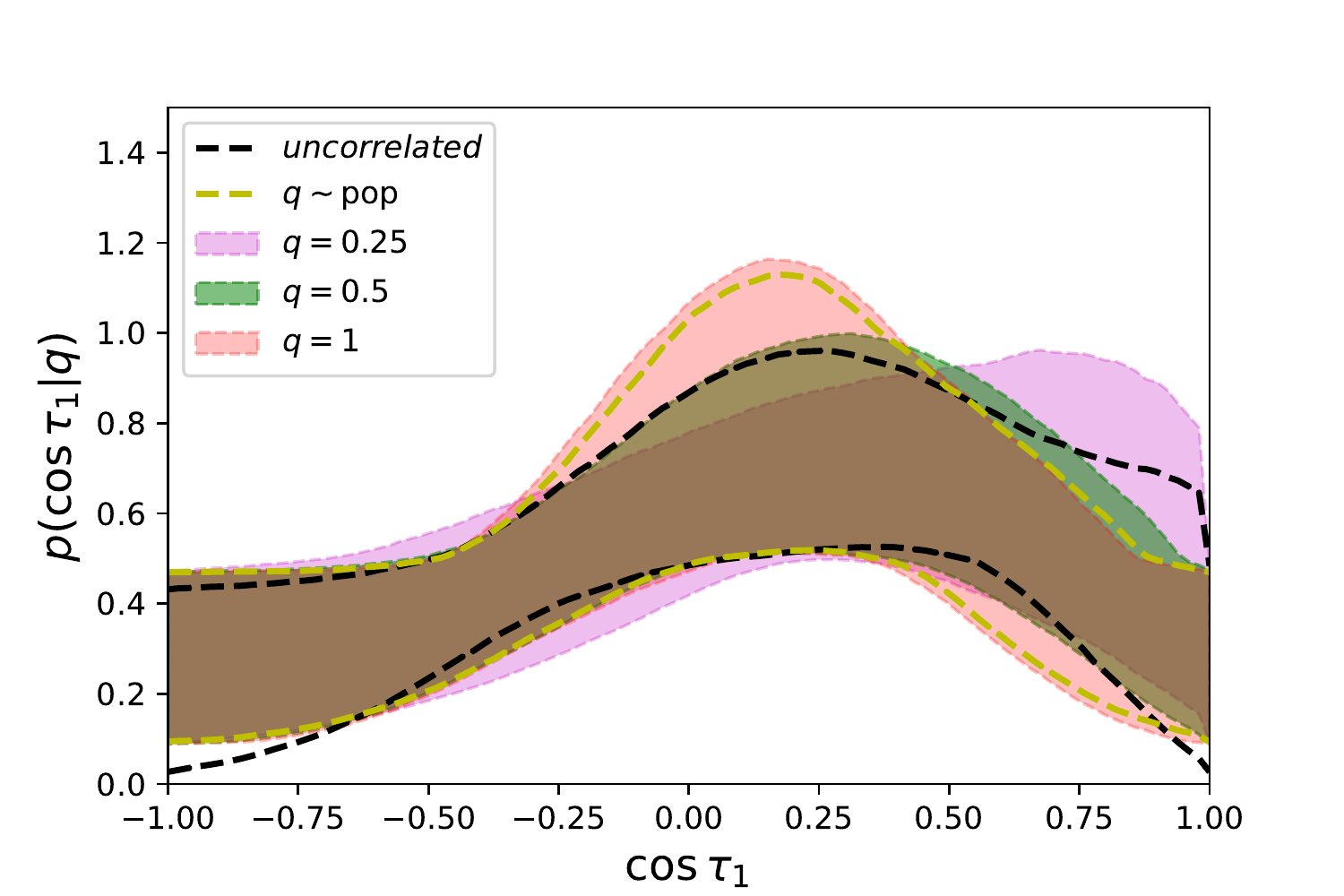}\\
\caption{Conditional posteriors for $\ct_1$ for the \isocbeta model, when $\ct$ is correlated with the mass ratio. Colored bands show the 90\% CI posterior for $\ct_1$ conditional on a fixed value of the mass ratio; yellow dashed lines enclose the 90\% CI obtained sampling the correlated parameter from its inferred astrophysical distribution; black dashed lines enclose the 90\% CI of the \isobeta model, which does not allow for correlations. The increased support at $+1$ as $q$ decreases is mostly prior-driven.}
\label{fig.iso_correlated_beta_costau}
\end{figure}

We consider the same possible correlations (i.e. values of $\kappa$ and $N$) described in the previous section. {As for the previous correlated model, we restrict the priors for $\alpha_b$ and $\beta_b$ to the non-negative domain to ensure that the overall $\alpha$ and $\beta$ parameters of the Beta distribution do not become negative, enforcing that only positive correlations can exist between the \ct distribution and $\kappa$.} 
For this model we find that the current dataset cannot significantly constrain the correlation parameters, even though we don't recover exactly the priors. For example, in Fig.~\ref{fig.IsoPlusCorrelatedBeta_q_BetaPars} we show the posterior and priors (thin dashed lines) for the parameters of the Beta distribution when we allow correlations with the mass ratio $q$. The parameters that enact the correlation, $\alpha_b$ and $\beta_b$ have wide posteriors, which however are not as similar as their prior as $\sigma_b$ was for the \isocgaus models, Fig.~\ref{fig.IsoPlusCorrelatedGaussian_m1_GaussianPars}. Figure~\ref{fig.iso_correlated_beta_q_beta_of_q} shows that the main impact of the measurement, relative to the prior, is to exclude large values of $\beta$. However, it is still the case that the overall trend in the 90\% CI of the Beta parameters are prior dominated. Just as for the \isocgaus models, this results in more support at $\ct\simeq +1$ for small masses, mass ratios or spins. Functionally, this happens because the parameters controlling the Beta component take smaller values at small values of the correlated parameter, and that moves the peak toward the edge of the domain---e.g. Fig.~\ref{fig.iso_correlated_beta_costau} for correlations with the mass ratio.
However, just as for the \isocgaus model, these trends are mostly a result of the prior and of the model. 

\clearpage
\section{Tukey window}\label{Sec.AppTukey}

We implement the Tukey window used in Eq.~\ref{Eq.IsoPlusTukey} as 

\begin{widetext}
\begin{align}
\mathcal{T}(x,T_{x0},T_k,T_r) &\propto
\begin{cases}
    0,\; & x< \rm{max}(-1,T_{x0}-T_k) \\
      \frac{1}{2} \left\{1+ \cos\left[\frac{\pi}{T_k T_r}\left(x-T_{x0}+T_k -T_k T_r \right)\right] \right\},&\; \mathrm{max}(-1,T_{x0}-T_k)  \leq x< T_{x0} -T_k (1-T_r)\\
      1,&\;  T_{x0} -T_k (1-T_r) \leq x < T_{x0} +T_k (1-T_r)\\
      \frac{1}{2} \left\{1+ \cos\left[\frac{\pi}{T_k T_r}\left(x-T_{x0}-T_k -T_k T_r \right)\right] \right\},&\;  T_{x0} +T_k (1-T_r) \leq x\leq \rm{min}(+1,T_{x0}+T_k) \\
      0,&\; x> \mathrm{min}(+1,T_{x0}+T_k)
    \end{cases}\,.
    \label{Eq.TukeyDef}
    \end{align}
\end{widetext}

\begin{figure}
\includegraphics[width=0.5\textwidth]{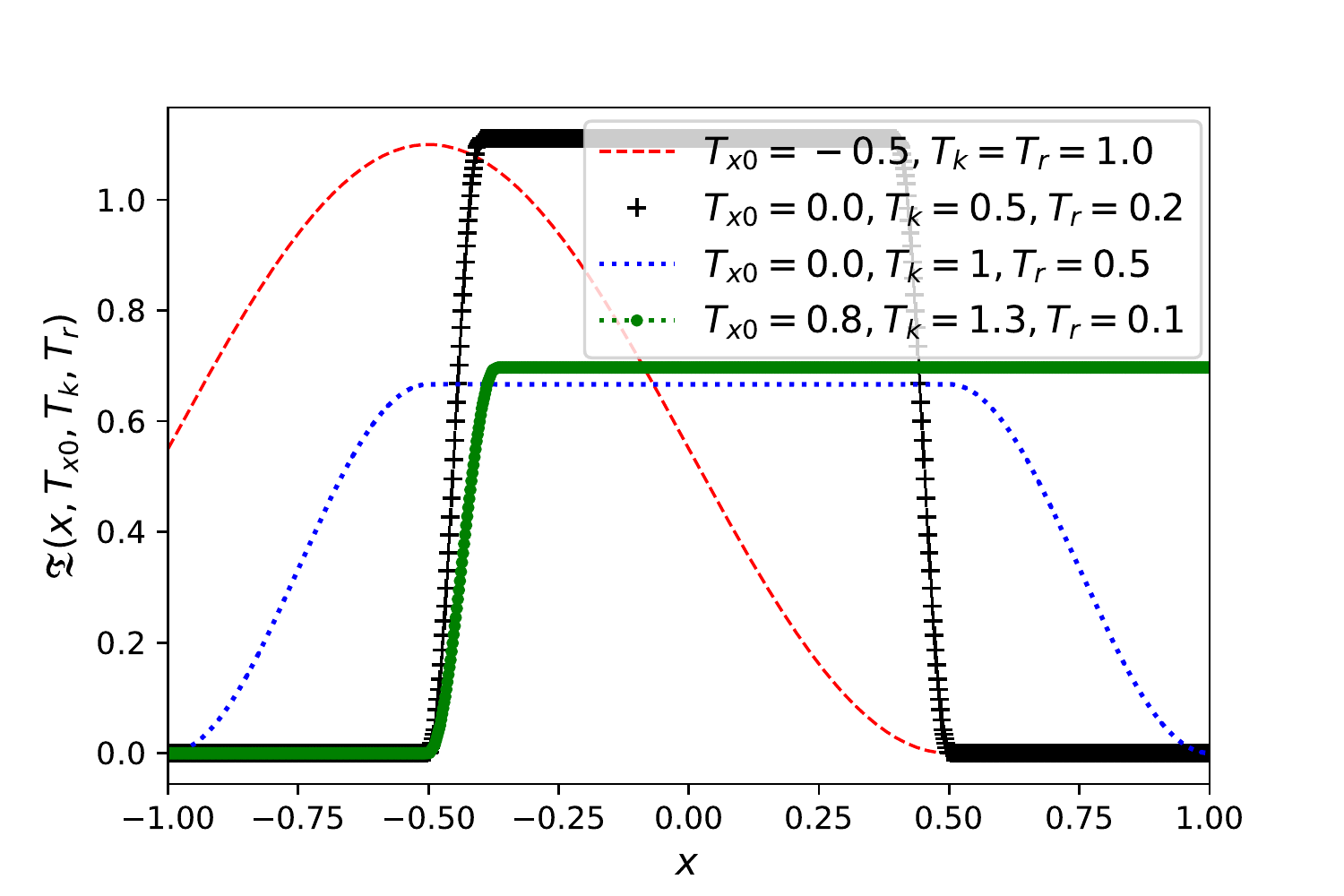}
\caption{
    Four examples of the distribution in Eq.~\ref{Eq.TukeyDef}.
}
\label{fig.Tukey_examples}
\end{figure}

This represents a Tukey window that is symmetric around  $T_{x0}$ and whose domain is $2 T_k$ wide. The parameter $T_r$ controls the shape of the window ($T_r=0$ gives a rectangular window while $T_r=1$ gives a cosine). The distribution is then truncated and normalized in the range $[-1,1]$. Figure~\ref{fig.Tukey_examples} shows four examples. Since the width, the shape, and the position can all be varied, this model is quite elastic and can latch onto both broad and narrow features. We highlight that in the default setting, we allow the uniform prior of $T_k$ to go up to 4, Tab~\ref{tab:priors}. This implies, that just as for the \isogaus model, there are parts of the parameter space where the non-isotropic component can be made very similar to, or indistinguishable from, the isotropic component. In this case, that happens when $T_k$ is large and $T_r$ is small. This distribution can also produce curves that ramp up from zero to a plateau, with various degrees of smoothness: the green thick line in Fig.~\ref{fig.Tukey_examples} is an example and---if $\tuk$ were zero---would produce a \ct distribution similar to the second row in Fig.~5 of~\citet{Callister:2022qwb}.

\clearpage
\section{Asymmetry $Y(\delta)$ for various values of $\delta$}\label{App.AllYs}

\begin{widetext}
\begin{figure}
\includegraphics[width=\textwidth]{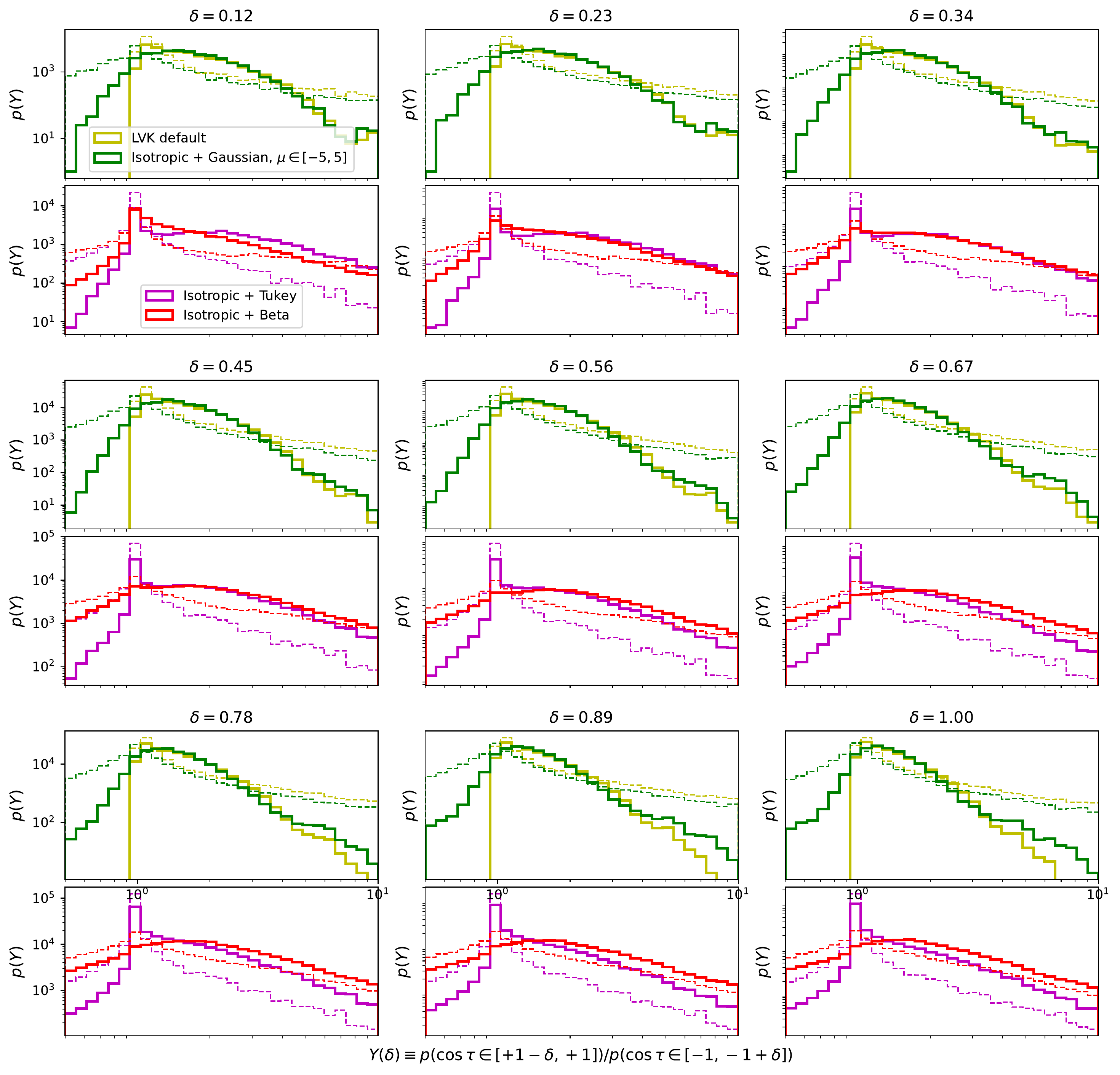}
\caption{Same as ~\ref{fig.Ys_comparison} but for various values of $\delta$.}
\end{figure}
\end{widetext}

\section{Tables}
Table~\ref{tab:Evidences} reports for all models (including those discussed later in other appendices) the number of spin parameters, the natural log of the Bayesian evidence and the natural log of the maximum likelihood point. All quantities are expressed as deltas relative to the reference \lvk model.  
Table~\ref{tab:priors} lists the priors used for the hyperparameters of all models. 
\clearpage

\begin{table}
\begin{tabular}{p{3.5cm} p{1.2cm}  p{1.2cm} p{1.5cm}}
\hline
   {Run} & $\ln$Evidence & $\ln$MaxL & \# spin pars \\
   \hline
   \hline
   \texttt{Isotropic} & $-0.8$& $-3.2$  & -2 \\
   \hline
   \lvk & \rm{ref} &  \rm{ref} & \rm{ref}\\
   \hline
   \isogaus \mbox{$\mu \in [-1,1]$} & $-0.3$ & $-0.4$ & +1 \\
  \hline
  \isogaus \mbox{$\mu \in [-5,5]$ }& $-0.4$ & $-0.2$ & +1 \\
  \hline
 \isobeta  & $+0.4$ & $+0.4$ & +1 \\
 \hline
  \isotuk  & $-0.2$ & $+0.7$ & +2 \\
 \hline
   \isocgaus ($m$) & $-0.4$ & $-0.6$ & +3 \\
  \hline
  \isocgaus ($q$) & $-0.5$ & $-0.1$ & +3 \\
  \hline
  \isocgaus ($\chi$) & $-0.6$ & $-0.4$ & +3 \\
  \hline
  \isocgaus ($M_{\rm{tot}}$) & $-0.7$ & $-0.1$ & +3 \\
  \hline
  \isocbeta ($m$) & $-0.4$ & $+0.0$ & +3 \\
  \hline
  \isocbeta ($q$) & $-0.1$ & $-0.3$ & +3 \\
  \hline
  \isocbeta ($\chi$) & $+0.3$ & $-0.6$ & +3 \\
  \hline
  \isocbeta ($M_{\rm{tot}}$) & $-0.3$ & $-0.7$ & +3 \\
 \hline
 \isogbeta  & $+0.3$ &$-0.6$ & +4 \\
 \hline
  \isoggaus \mbox{$(-A=B=1)$}& $-0.1$ &$-0.3$ & +4 \\
   \hline
 \isogtuk  & $+0.1$ &$+0.3$ & +5 \\
 \hline
\isoggaus & $+0.6$ &$+0.2$ & +6 \\
  \hline
  \hline
\end{tabular}
\caption{Bayesian evidence, maximum log likelihood value an number of parameters for the \ct models (relative to the reference LVK model of Eq.~\ref{Eq.LVKModel}). With our settings, the evidences carry a statistical uncertainty of $\pm 0.15$.
Additional uncertainties in the log likelihood -- and hence evidence -- arise from the numerical evaluation of the integral in Eq.~\ref{eq.Sampling}. \citet{Golomb:2022bon} estimates that to be roughly $\Delta\log  \mathcal{L}\pm 1$ when using the publicly released LVK injection sets to estimate selection effects. We also note that nested sampling algorithms do not aim at finding the highest likelihood point, so it is possible that for nested models (e.g. \lvk and \isogaus) the broader model finds a slightly lower maximum likelihood point. Given these uncertainties, the only reliable -- yet unsurprising -- conclusions one may draw is that a purely isotropic model yields the worst match to the data.
}
\label{tab:Evidences}
\end{table}

\begin{table}
\centering
\begin{tabular}{l || c }
\hline 
\hline 
\multicolumn{2}{c}{\lvk - Eq.~\ref{Eq.LVKModel}}\\
\hline 
$\sigma$ & \Un~(0.1,4) \\
\gau & \Un(0,1) \\
\hline
\multicolumn{2}{c}{\isogaus - Eq.~\ref{Eq.IsoPlusGaussian}}\\
\hline
$\mu$ & \Un(-5,5) or \Un(-1,1) \\
$\sigma$ & \Un~(0.1,4) \\
\gau & \Un(0,1) \\
\hline
\multicolumn{2}{c}{\isobeta - Eq.~\ref{Eq.IsoPlusBeta}}\\
\hline
$\alpha$& \Un(0.05,5) \\
$\beta$& \Un(0.05,5) \\
\bet & \Un(0,1) \\
\hline
\multicolumn{2}{c}{\isotuk - Eq.~\ref{Eq.IsoPlusTukey}}\\
\hline
$T_{x0}$& \Un(-1,1) \\
$T_{r}$& \Un(0.01,1) \\
$T_{k}$& \Un(0.1,4) \\
\tuk & \Un(0,1) \\
\hline
\multicolumn{2}{c}{\isogbeta  - Eq.~\ref{Eq.IsoPlusGaussianPlusBeta}}\\
\hline
$\alpha$& \Un(1,20) \\
$\beta$& \Un(1,20) \\
$\mu$ & \Un(0.9,5) \\
$\sigma$ & \Un~(0.1,5) \\
\gau,\bet & \Un(0,1), $\gau+\bet\leq 1$ \\
\hline
\multicolumn{2}{c}{\isoggaus - Eq.~\ref{Eq.IsoPlusGaussianPlusGaussian}}\\
\hline
$\mu_L$ & \Un(-1,1), $A<\mu_L<B$ \\
$\sigma_L$ & \Un~(0.1,4) \\
$\mu_R$ & \Un(0.9,5) \\
$\sigma_R$ & \Un~(0.1,5) \\
$A$ & \Un(-1,0.1) \\
$B$ & \Un(0.2,0.9) \\
$\gau_L$,$\gau_R$ & \Un(0,1), $\gau_L+\gau_R\leq 1$ \\
\hline
\multicolumn{2}{c}{\isogtuk - Eq.~\ref{Eq.IsoPlusGaussianPlusTukey}}\\
\hline
$T_{x0}$& \Un(-1,1) \\
$T_{r}$& \Un(0.01,1) \\
$T_{k}$& \Un(0.1,4) \\
$\mu_R$ & \Un(0.9,5) \\
$\sigma_R$ & \Un~(0.1,5) \\
$\gau$,$\tuk$ & \Un(0,1), $\gau+\tuk\leq 1$ \\
\hline
\multicolumn{2}{c}{\isocgaus - Eq.~\ref{Eq.IsoPlusCorrelatedGaussian}}\\
\hline
$\mu_a$ & \Un(-5,5)  \\
$\mu_b$ & \Un(-5,5)  \\
$\sigma_a$ & \Un~(0.1,4) \\
$\sigma_b$ & \Un~(0,10) \\
\gau & \Un(0,1) \\
\hline
\multicolumn{2}{c}{\isocbeta - Eq.~\ref{Eq.IsoPlusCorrelatedBeta}}\\
\hline
$\alpha_a$ & \Un(0.05,5)  \\
$\alpha_b$ & \Un(0,10)  \\
$\beta_a$ & \Un(0.05,5)  \\
$\beta_b$ & \Un(0,10)  \\
\bet & \Un(0,1) \\
\hline
\hline
\end{tabular}
\caption{The priors used for the hyper parameters of the tilt models. All variables are dimensionless. }
\label{tab:priors}
\end{table}

\end{document}